\documentclass[a4paper,fleqn,usenatbib]{mnras}

\usepackage{amsmath}	
\usepackage{newtxtext,newtxmath}

\usepackage[T1]{fontenc}
\usepackage{ae,aecompl}

\usepackage{graphicx}	

\usepackage{amssymb}	

\usepackage{threeparttable}
\usepackage{color}
\usepackage{bm}
\usepackage{epsfig}
\usepackage{wrapfig}
\usepackage{multicol}
\usepackage{exscale}
\usepackage{enumitem}

\newcommand{\Msun}{\mbox{$\rm M_{\odot}$}} 

\newcommand{\Lsun}{\>{\rm L_{\odot}}} 
\newcommand{\kpc}{\>{\rm kpc}}

\title[Quantifying the Origin and Distribution of Intracluster Light in a Fornax-like Cluster]{Quantifying the Origin and Distribution of Intracluster Light}

\author[K.A.Harris et al.]{Kathryn A. Harris$^{1}$\thanks{E-mail:
    kateharris142@gmail.com}, Victor P. Debattista$^{2}$, Fabio
  Governato$^{3}$, \newauthor Benjamin B. Thompson$^{2}$, Adam J.
  Clarke$^{2}$, Thomas Quinn$^{3}$, Beth Willman$^{4}$, \newauthor
  Andrew Benson$^{5}$, Duncan Farrah$^{1}$, Eric W. Peng$^{6,7}$,
  Rachel Elliott$^{1}$, Sara Petty$^{1}$ \\
  $^{1}$ Department of Physics, Virginia Tech, Blacksburg, VA 24061, USA \\
  $^{2}$ Jeremiah Horrocks Institute, University of Central Lancashire, Preston PR1 2HE \\
  $^{3}$ Astronomy Department, University of Washington, WA 98195, USA \\
  $^{4}$ Haverford College, Haverford, PA 19041, USA \\
  $^{5}$ The Observatories of the Carnegie Institution for Science, 813 Santa Barbara St, Pasadena, CA 91101, USA \\
  $^{6}$ Department of Astronomy, Peking University, Beijing 100871, China  \\
  $^{7}$ Kavli Institute for Astronomy and Astrophysics, Peking University, Beijing 100871, China \\
}

\date{Accepted XXX. Received YYY; in original form ZZZ}

\pubyear{2015}

\begin{document}
\label{firstpage}
\pagerange{\pageref{firstpage}--\pageref{lastpage}}
\maketitle

\begin{abstract}

Using a cosmological $N$-body simulation, we investigate the origin
and distribution of stars in the intracluster light (ICL) of a
Fornax-like cluster.  In a dark matter only simulation we
  identify a halo which, at $z=0$, has $M_{200} \simeq 4.1 \times
  10^{13}\Msun$ and $r_{200} = 700\kpc$, and replace infalling subhalos
  with models that include spheroid and disc components.  As they fall into the
  cluster, the stars in some of these galaxies are stripped from their hosts, and form
  the ICL. We consider the separate contributions to the ICL from
stars which originate in the haloes and the discs of the galaxies. We
find that disc ICL stars are more centrally concentrated than halo ICL
stars. The
majority of the disc ICL stars are associated with one initially 
disc-dominated galaxy that falls to the centre of the cluster and is 
heavily disrupted, producing part of the cD galaxy. At radial distances greater than 200kpc, well beyond the stellar
envelope of the cD galaxy, stars formerly from the stellar
haloes of galaxies dominate the ICL.  Therefore at large distances,
the ICL population is dominated by older stars.
  
\end{abstract}

\begin{keywords}{cosmology: theory --- methods: numerical --- galaxies: clusters: general
--- galaxies: clusters: intracluster medium} \end{keywords}

  
\section{Introduction}
\label{sec:intro}

Intracluster light (ICL), discovered by \citet{Zwicky1951}, is light
from stars within a galaxy cluster which are not bound to a
galaxy. This light, which has been observed in clusters up to 
$z = 0.8$ \citep{Guennou2012}, constitutes between 2\% and 50\% of a
cluster's total light \citep{Arnaboldi2004, Feldmeier2004, Lin2004,
  Zibetti2008, McGee2010, Toledo2011}.  The origin of these stars is
thought to be the cluster galaxies themselves; the stars being stripped from the galaxies as they orbit within the cluster
\citep[e.g.][]{White1976, Merritt1984, Moore1996, CalcaneoRoldan2000,
  Arnaboldi2004, Zibetti2005, Conroy2007}.  Indeed the network of
tidal features observed in the core of the Virgo cluster supports the
view that stripped stars are a major contributor to the ICL
\citep{Mihos2005}.  Nonetheless, an ICL component formed {\it in situ} out
of gas has also been suggested.  For instance, the simulations of
\citet{Puchwein2010}, which include AGN feedback, find that as much as
$30\%$ of the ICL may have formed {\it in situ}. However, there may be an upper limit to how much ICL light in situ star formation can produce.  While gas being stripped from
infalling galaxies is sometimes seen to be forming stars
\citep[e.g.][]{Sun2010, Zhang2011}, stellar population synthesis of
stacked spectra of the ICL give an upper limit to a younger ICL
population of $\sim 1\%$ \citep{Melnick2012}.

The ICL is an important diagnostic of processes such as the enrichment
of the intracluster medium \citep{Sivanandam2009, Cora2006,Lin2004}
which affects cooling flows \citep[][]{Peterson2003},
the evolution of baryonic substructures \citep{Arnaboldi2010}, and the
baryon budget in clusters and groups \citep{Gonzalez2005,
  Gonzalez2007}.  The ICL has also been used to trace the structure of
dark matter: \citet{Jee2010} used the ICL and weak lensing to trace a
ring-like dark matter structure near the centre of the cluster
CL0024+17. \citet{Giallongo2014} compared observations of the ICL to a
dark matter model and concluded that it can be used to probe the dark
matter distribution.

The ICL fraction depends on galaxy richness, increasing from $<2\%$
for loose groups \citep[e.g.][]{Feldmeier2003, Feldmeier2004} to
$>20\%$ for rich clusters \citep[e.g.][]{Feldmeier2002,Lin2004}. This effect is also seen in some simulations \citep{Rudick2011}.
Other simulations however have found a fairly constant ICL fraction with cluster richness
\citep[e.g.][]{Puchwein2010}.  The semi-analytic model of
\citet{Purcell2007} finds an ICL fraction rising rapidly to $\sim
20\%$ in haloes with mass up to $10^{13} ~\Msun$, but that this rise becomes much less dramatic at higher masses, reaching only $\sim 30\%$ at $10^{15} ~\Msun$.
The semi-analytic model of \citet{Contini2014} instead produces no
halo mass dependence.

Although the ICL can make up to half of the cluster light, it is much
more diffuse than the light from the galaxies
\citep[e.g.][]{Feldmeier2002, Guennou2012}, making it difficult to
detect.  Instead, the ICL is often studied using discrete, bright
tracers such as planetary nebulae \citep[e.g.][]{Arnaboldi1996,
  Arnaboldi2005, Aguerri2005, CastroRodriguez2009, Mihos2009}, red
giants \citep[e.g.][]{Ferguson1998,Durrell2002, Palladino2012},
globular clusters \citep[e.g.][]{Williams2007, Peng2011,Durrell2014},
novae \citep[e.g.][]{Neill2005, Shara2006} and supernovae
\citep[e.g.][]{Smith1981, GalYam2003, McGee2010, Sand2011}. These
methods however require assumptions about the light from the
underlying stellar population in order to convert between the observed
number of tracers and the total stellar mass in the ICL.

Numerical simulations have been used to study the ICL. These include
the creation and evolution of the ICL \citep{Rudick2006, Puchwein2010,
  Cooper2015}, the kinematics of unbound stars
\citep[e.g.][]{Murante2004, Willman2004,Dolag2010}, the ICL fraction
\citep[e.g.][]{Murante2004, Willman2004, Rudick2011, Martel2012}, ICL
substructures
\citep[e.g.][]{CalcaneoRoldan2000, Murante2004, Rudick2011}, and the
radial distribution of the ICL \citep[e.g.][]{Napolitano2003,
  Rudick2009, Guennou2012}. A common feature of most of these
simulations is that they focus on high-mass clusters, generally
$>10^{14} ~\Msun$. However, high mass clusters are atypical, with most
galaxies residing in lower mass associations, such as loose groups or
clusters.  These lower mass clusters have
several potentially key density and kinematic differences with their
high-mass cousins, \citep[e.g.][]{Forman1990}. For example, the Fornax
cluster, though less massive than the Virgo cluster
\citep[$\sim10^{13} - 10^{14} ~\Msun$, as opposed to Virgo's
  $\sim10^{15} ~\Msun$][] {Ikebe1992, Drinkwater2001, Nasonova2011,
  Fouque2001, Lee2015}, has approximately three times the density of galaxies
within its core compared to Virgo \citep{Davies2013}, and a lower
velocity dispersion of the giant galaxies ($\sim 370$ km s$^{-1}$
compared to $\sim 1000$ km
s$^{-1}$ \citep{Ftaclas1984, Binggeli1993, Drinkwater2000, Drinkwater2001,Kim2014}).  Fornax also
appears to be more dynamically evolved than Virgo
\citep{Churazov2008}.  Because of these differences, coupled with the
way the ICL may be formed (via tidal interactions and stripping from
galaxies), we may expect different ICL compositions and distributions in lower mass
clusters.

This paper investigates the distribution and origin of ICL stars in a
cluster comparable to the Fornax cluster using an $N$-body
simulation. This simulated cluster is of a lower mass than has been
studied in previous ICL papers, with a total stellar mass of $\sim
10^{11} ~\Msun$ and dark matter mass of $\sim 10^{13} ~\Msun$. We
investigate whether ICL stars originate primarily from the discs or
the haloes of progenitor galaxies. Separating the ICL into stars
which originated from the disc (younger) and halo (older) of galaxies
allows us to further investigate the idea that the ICL stars may be
older than the stellar populations in surviving galaxies
\citep[e.g.][]{Murante2004}.  Section 2 describes the simulation used
and the formation of part of the cD galaxy.  In Section 3,
methods used to image and identify the ICL are described. The results
are presented in Section 4 including the radial distribution of the
ICL and its components, the ICL luminosity, and the stellar age of the
ICL. Section 5 discusses these results and presents our conclusions.
In the Appendix we present the individual galaxies and their
  orbits within the cluster.


\section{$N$-body Methods}
\label{sect:sim}

We explore the ICL via a cosmological simulation using
\textsc{pkdgrav} \citep{Stadel2001} of 18 galaxies within a Fornax
cluster-like environment. We start with a dark matter (DM) only
simulation, evolved in a WMAP \citep{Spergel2003} $\Lambda$CDM
cosmology with $\Omega_0 = 0.3$, $\Omega_{\Lambda} = 0.7$ and $H_0 =
70$ km s$^{-1}$ Mpc$^{-1}$.  The initial simulation has a low
resolution of $36^3$ particles in a cube of size $70 h^{-1}$ Mpc.  We
then identify a cluster with a virial radius $r_{200} \simeq 700$ kpc
and virial mass $M_{200} \simeq 4.1\times 10^{13} \Msun$.  These
properties make the cluster similar to the Fornax cluster. We then use
the zoom-in technique \citep{Katz1993} to re-simulate at high
resolution the formation of the Fornax-sized halo. The resolution
  of the base simulation is refined in four nested steps by factors
of $2^3$, $2^3$, $4^3$ and $3^3$ centred on the cluster.  At the
highest resolution this gives a total of 15,492,788 dark matter
particles.  The particle masses range from $7.9 \times 10^6 ~\Msun$ to
$8.7\times10^{11} ~\Msun$, with corresponding particle softenings
varying from $0.13~\kpc$ to $6.2~\kpc$.

At the end of this process we identified all haloes in the mass range
$8.6 \times 10^{10} ~\Msun \leq ~M_{200} \leq 5.2 \times 10^{12} ~\Msun$
entering the cluster between redshift $z = 1.65$ and $0.13$ (before
$z = 1.65$ the cluster is too chaotic to allow easy replacement).
Before they enter, we replace 18 of these haloes with full galaxy
models.  We exclude only 2 haloes from this replacement because they
are strongly interacting as they enter the cluster.  Because strong
interactions eject stars to large radii \citep{Hilz2012, Hilz2013}
where they are more easily stripped in a cluster environment,
excluding these haloes will lead to an underestimate of the ICL.
Material that fell into the cluster earlier than $z = 1.6$ would
probably have ended up in the central cD galaxy \citep{Diemand2005}.
Thus the stellar content of the central object in our simulation is
underestimated.
Our replacement mass cutoff for $M_{200} > 8.6 \times 10^{10} ~\Msun$
roughly corresponds to a stellar mass cut of $\sim 10^9 ~\Msun$,
equivalent to $< 0.1 ~L_*$ at $z=0$ \citet{DeMaio2015} show the ICL is
likely to be dominated by stripping from $>0.2L_*$ galaxies rather
than stripping or disruption of dwarfs or mergers with the central
galaxy.  Thus the mass limit is unlikely to have introduced
significant biases in the ICL in our simulation.

We then re-run the simulation replacing these haloes by full
bulge+disc+DM models as they fall in.  This
procedure assumes that the star formation of each field galaxy
proceeded normally and was then interrupted by ram pressure stripping
of the gas upon entering the dense cluster environment
\citep[e.g.][and references therein]{Peng2010, Taranu2014, Tal2014}.

Semi-analytic models are very useful for populating systems growing
hierarchically \cite[e.g.][]{Moster2014}.  For our model galaxies to
resemble observed galaxies, we select their parameters guided by the
semi-analytic model catalogues of \citet{Cole2000}.  These
semi-analytic models employed a cosmology with $H_0 = 69.7$ km
s$^{-1}$ Mpc$^{-1}$, $\Omega_0 = 0.3$, $\Omega_b = 0.02$, and
$\Omega_{\Lambda} = 0.70$.  Though this semi-analytic model has a
lower value of $\Omega_b$ than the concordance cosmology
\citep{Bennett2013}, the parameters of the semi-analytic model were
tuned to match the observed galaxy properties.  As such, sampling
galaxy properties from this semi-analytic model is a viable way of
selecting structural parameters for galaxies that are replaced.  The
galaxies in the catalogues were produced at 20 epochs, equally spaced
in time between $z=0$ and $z=6$.  Each redshift catalogue contained
between $\sim 7100$ and $\sim 8200$ galaxies.  

The \citet{Cole2000} catalogues were modelled in a volume of $10^5
~h^3$ Mpc$^{-3}$ and with masses spanning $5\times10^{9}$ to $1\times
10^{15} ~h^{-1} ~\Msun$.  The galaxy catalogues generated in this way
contain information on the structural properties of the galaxies
(masses, sizes, age and metallicity of both disc and halo components)
and dark matter haloes (mass, virial velocity, concentration and spin
parameter).  In selecting the best galaxy model for each halo, we used
the semi-analytic catalogues, choosing the halo which best matches the
target $M_{200}$ and $V_{200}$ at the infall redshift; however we
discard matches where the bulge-to-disc mass ratio, $B/D < 1/9$, which
we arbitrarily chose so the spheroid is well populated.

The initial model galaxies were generated using the method of
\citet{Springel1999}.  The models consist of exponential discs,
Hernquist bulges \citep{hernqu_90} and NFW DM haloes
\citep*{nav_etal_97_nfw}.  In all cases, the vertical profile of the
disc is $\mathrm{sech}^2 z/z_d$, with ratio of scale-height, $z_d$, to
scale-length, $R_d$, set to $z_d/R_d = 0.1$.  Table \ref{tab:galaxies}
lists the initial conditions of the galaxies, including the redshift
at which each dark matter halo is replaced, the halo mass, the
circular velocity at $r_{200}$, the stellar mass, the disc-scale
length, the bulge-to-disc mass ratio, and the ratio of the bulge
effective radius to disc scale-length, $R_e/R_d$.  We set stellar
particle softening to $\epsilon = 0.03 R_d$, which ends up
corresponding to $\epsilon = 34$ pc to $198$ pc.  We use a larger
softening, in the range $\epsilon = 0.57$ kpc to $5.1$ kpc, for dark
matter particles.  In all cases the discs consist of 300,000
particles, while the number of dark matter particles varies between
301,719 and 944,783.

The Hernquist bulges formally extend to large radii.  Rather than
truncating these bulges at some finite radius, we treat the star
particles at large radius as the stellar halo particles.
\citet{deJong2009} found that bulge and stellar halo density profiles
in nearby disc galaxies join smoothly, so it is reasonable to
identify the model's outer bulge as the stellar halo, at least from
the density profile point of view.  However, the actual ratio of
  stellar halo to disc stars remains model-dependent because the
assumed functional form of both the bulge (Hernquist bulge) and
the disc (single exponential) are somewhat arbitrary at large
radii and certainly difficult to constrain at high redshift.  Thus,
the ratio of disc to halo stars in the disc outskirts is poorly
constrained.  Because in general disc density profiles tend to
truncate at large radii \citep[e.g.][]{vanderKruit1979,
  vanderKruit1987, Pohlen2000, Pohlen2002, Pohlen2006, Erwin2005}, we
are very likely over-representing the disc contribution to the ICL.
The surface brightness profiles for each galaxy as they enter
the cluster are presented in Appendix \ref{Appendix 1}.

We insert each model galaxy before it crosses the virial radius of the
cluster, in order to give the models sufficient time to relax,
  $>100$ Myr, since the initial conditions generated using the method
of \citet{Springel1999} are not in perfect equilibrium
\citep{Kazantzidis2004}.

\begin{figure} 
\begin{center}
\includegraphics[width=0.35\textwidth,angle=0]{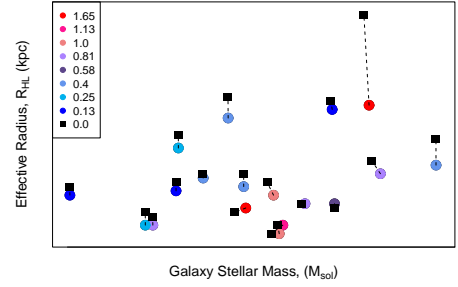}
\end{center}
\caption{The size-mass relation of our galaxies when inserted into the
  simulation (circles coloured for redshift) and at $z=0$ (black
  squares). The dashed lines show the evolution of the mass and
  stellar size for each galaxy.  }
\label{fig:sizemass} 
\end{figure}

We compare our galaxies to the observational data of
\citet{Williams2010} and the simulations of \citet{Laporte2013} out to
redshifts of $z \sim 2$.  Fig. \ref{fig:sizemass} shows the size-mass
relation of our 18 galaxies when they are first inserted into the
simulation, with redshifts between 1.65 and 0.13 and at $z=0$. At
$z=0$, we define the half light radius, $R_{HL}$, of the galaxies
using only stars within our radial cut at 50 kpc (corresponding to the
radial cut we use to separate galaxies from ICL described in \S
\ref{sect:ICL}).  Across the redshift range $0<z<2$, our selected
galaxies lie within the region of observed galaxies from
\citet{Newman2012}. The galaxies are slightly less massive than those
simulated by \citet{Laporte2013}. At redshift slightly higher than the
redshift of our first replaced galaxy, the galaxies lie within
1$\sigma$ of the relation of \citet{Williams2010}. Therefore our
selection of galaxy parameters is realistic.  The overall trend
revealed by Fig. \ref{fig:sizemass} is for $R_{HL}$ to increase.
Galaxies of stellar mass $M_* > 5\times 10^9 ~\Msun$ experience most
of the stripping, while lower mass galaxies are not as heavily
stripped.  Therefore most of the ICL comes from massive galaxies.

\begin{figure} 
\begin{center}
\includegraphics[width=0.5\textwidth]{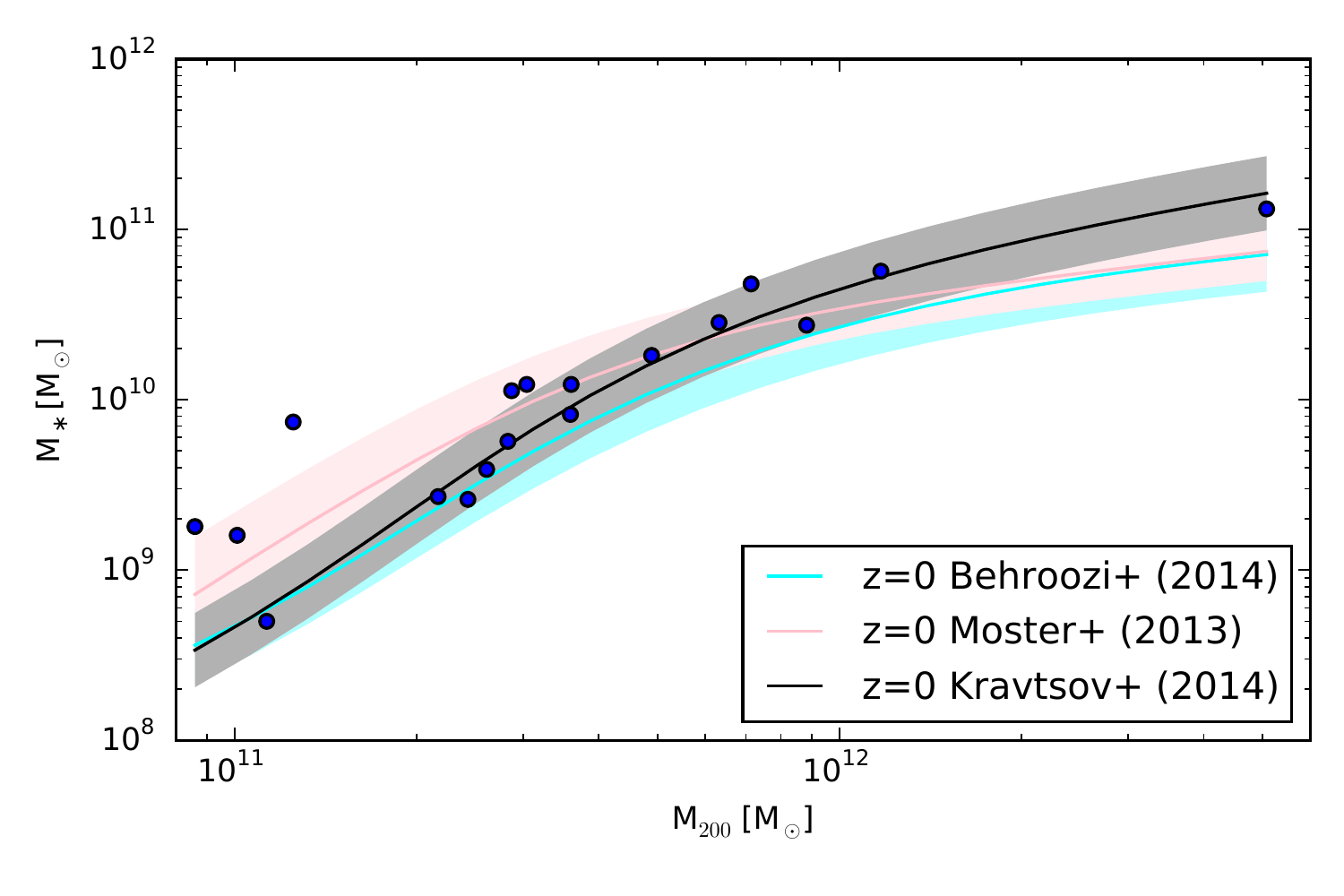}
\end{center}
\caption{Stellar mass versus halo mass  for each resimulated
  galaxy at the time at which they are introduced (blue points).
  Assuming these masses do not change if the models were evolved in
  isolation, we compare with the abundance matching relations from
different studies as indicated. The different shaded regions
correspond to the uncertainties in each of the relations.}
\label{fig:massmass} 
\end{figure}

As a final check of the model galaxy parameters, we plot in Fig.
\ref{fig:massmass}  the stellar mass versus halo mass for each
  resimulated galaxy at the time at which we introduce it into our
  N-body simulation.  If the galaxies were evolved in isolation,
  rather than falling into the cluster, we do not believe either of
  these quantities would change significantly between that time and
  $z=0$. We therefore compare these initial conditions directly to the
  $z=0$ abundance matching relations of \citet{Behroozi2013},
\citet{Moster2013}, and \citet{Kravtsov2014}.  Our model galaxies are
in good agreement with these relations, particularly with the relation
of \citet{Kravtsov2014}.

\begin{table*} 
\begin{centering} 
\caption{The model galaxies in the simulation.} 
\small{ 
\begin{tabular}{c|ccccccccc}
\hline 
Galaxy & $z_i^{a}$ & $M_{200}^{b}$  & $V_{200}^{c}$  & $M_*^{d}$   & $R_d^{e}$ & $B/D^{f}$ & $R_e/R_d^{g}$  &  $R_{HL}^{h}$  & $M_V^{i}$\\ 
       &       & $[10^{10} \Msun]$ & [kms$^{-1}$] & $[10^{10} \Msun]$ & [kpc]    &       &            &  [kpc]  \\ \hline
G01    & 1.65  & 71.4      & 180    & 4.79     & 2.5    & 0.12  & 0.13 &  3.9 & -19.4 \\
G02    & 1.65  & 12.5      & 106    & 0.74     & 1.1    & 0.27  & 0.17 &  1.5 & -17.4 \\
G03    & 1.13  & 30.4      & 136    & 1.23     & 0.7    & 0.15  & 0.25 &  1.1 & -18.0 \\
G04    & 1.00  & 36.0      & 135    & 1.23     & 0.7    & 0.95  & 0.41 &  0.9 & -17.9 \\
G05    & 1.00  & 28.7      & 127    & 1.13     & 1.7    & 1.11  & 0.31 &  1.8 & -17.8 \\
G06    & 0.81  & 48.9      & 144    & 1.82     & 1.1    & 0.18  & 0.12 &  1.6 & -18.3 \\
G07    & 0.81  & 8.6       &  82    & 0.18     & 0.6    & 0.12  & 0.41 &  1.1 & -15.8 \\
G08    & 0.81  & 117.0     & 190    & 5.69     & 2.5    & 6.84  & 0.44 &  2.3 & -19.5 \\
G09    & 0.58  & 63.2      & 150    & 2.84     & 1.1    & 0.19  & 0.21 &  1.6 & -18.8 \\
G10    & 0.40  & 28.3      & 112    & 0.57     & 2.3    & 0.13  & 0.17 &  3.6 & -17.1 \\
G11    & 0.40  & 35.9      & 121    & 0.82     & 1.3    & 0.13  & 0.13 &  2.0 & -17.3 \\
G12    & 0.40  & 26.1      & 107    & 0.39     & 1.4    & 0.12  & 0.26 &  2.2 & -16.7 \\
G13    & 0.40  & 508.0     & 278    & 13.2     & 2.1    & 0.52  & 0.14 &  2.5 & -20.5 \\
G14    & 0.25  & 21.7      &  95    & 0.27     & 1.9    & 0.17  & 0.26 &  2.9 & -16.3 \\
G15    & 0.25  & 10.1      &  76    & 0.16     & 1.5    & 1.29  & 0.14 &  1.1 & -15.7 \\
G16    & 0.13  & 88.2      & 148    & 2.74     & 2.1    & 0.84  & 1.16 &  3.8 & -18.8 \\
G17    & 0.13  & 11.3      &  72    & 0.05     & 1.1    & 0.57  & 0.76 &  1.8 & -14.4 \\
G18    & 0.13  & 24.3      &  96    & 0.26     & 1.1    & 0.15  & 0.54 &  1.9 & -16.2 \\ \hline 
\end{tabular}
\begin{tablenotes}
        \item $^a$ Redshift at which the halo is replaced with a full model \\
        \item $^b$ Halo viral mass\\
        \item $^c$ Viral velocity\\
        \item $^d$ Stellar mass\\
        \item $^e$ Disc scale-length\\
        \item $^f$ Bulge-to-disc mass ratio\\
        \item $^g$ Ratio of the bulge effective-radius to disc scale-length\\
        \item $^h$ Half light radius of the galaxy when placed in the simulation.\\
        \item $^i$ Absolute magnitude in the $V$-band. \\
\end{tablenotes}
}
\label{tab:galaxies} 
\end{centering} 
\end{table*}

\subsection{Tests of the simulation}
\label{ssect:tests}

\begin{figure} 
\begin{center}
\includegraphics[width=0.35\textwidth,angle=-90]{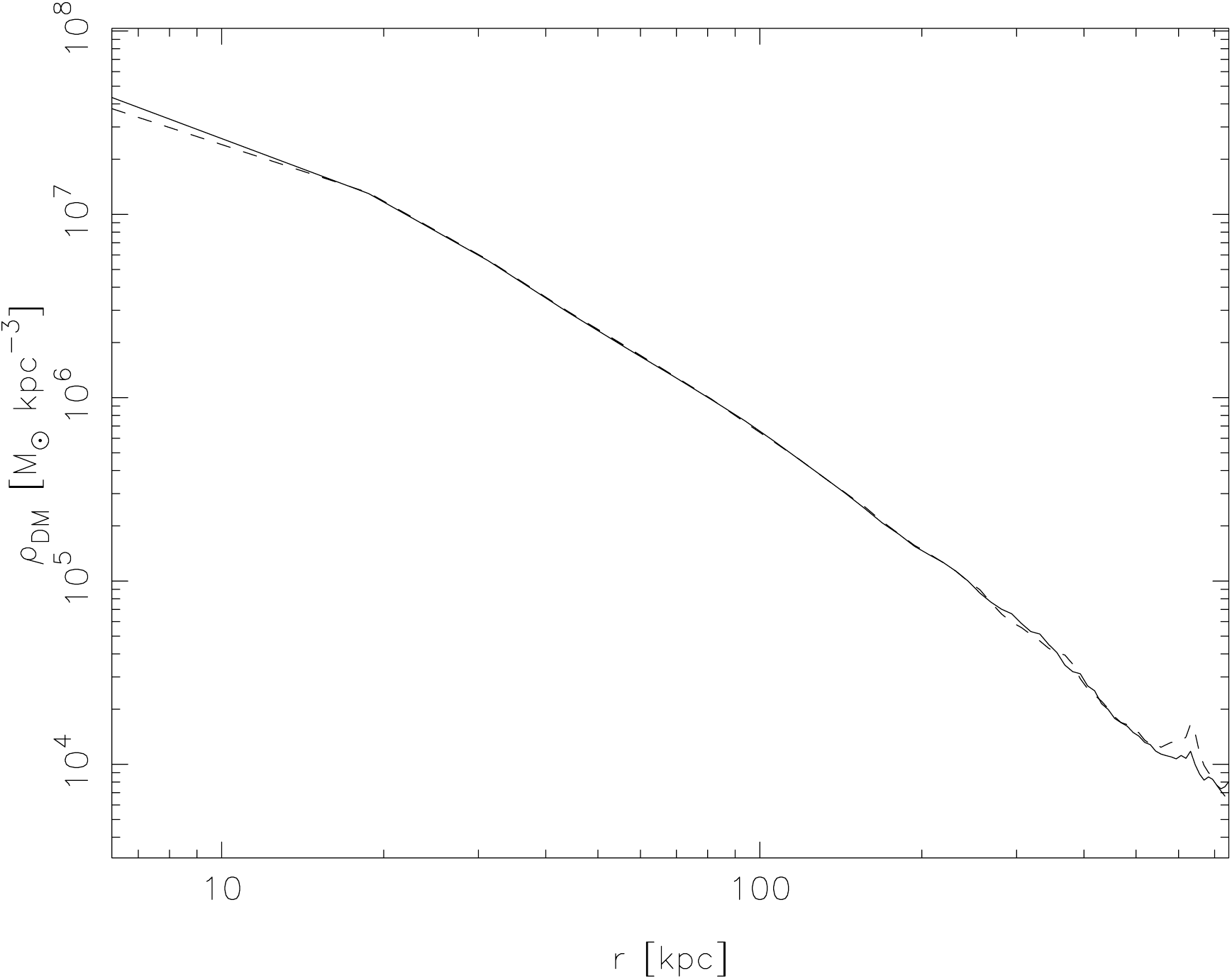}
\end{center}
\caption{The full dark matter density profile in the dark matter only
  simulation (solid line) and in our resimulation with infalling haloes
  replaced (dashed line).  }
\label{fig:dmprofiles} 
\end{figure}

We verified that our replacement of infalling dark matter haloes does
not alter the overall structure of the cluster substantially.  Fig.
\ref{fig:dmprofiles} shows the density profile of the dark matter
component in the dark-matter-only simulation and in our resimulation
with galaxy replacements.  It is apparent that our procedure
does not alter the cluster.

\begin{figure} 
\begin{center}
\includegraphics[width=0.35\textwidth,angle=-90]{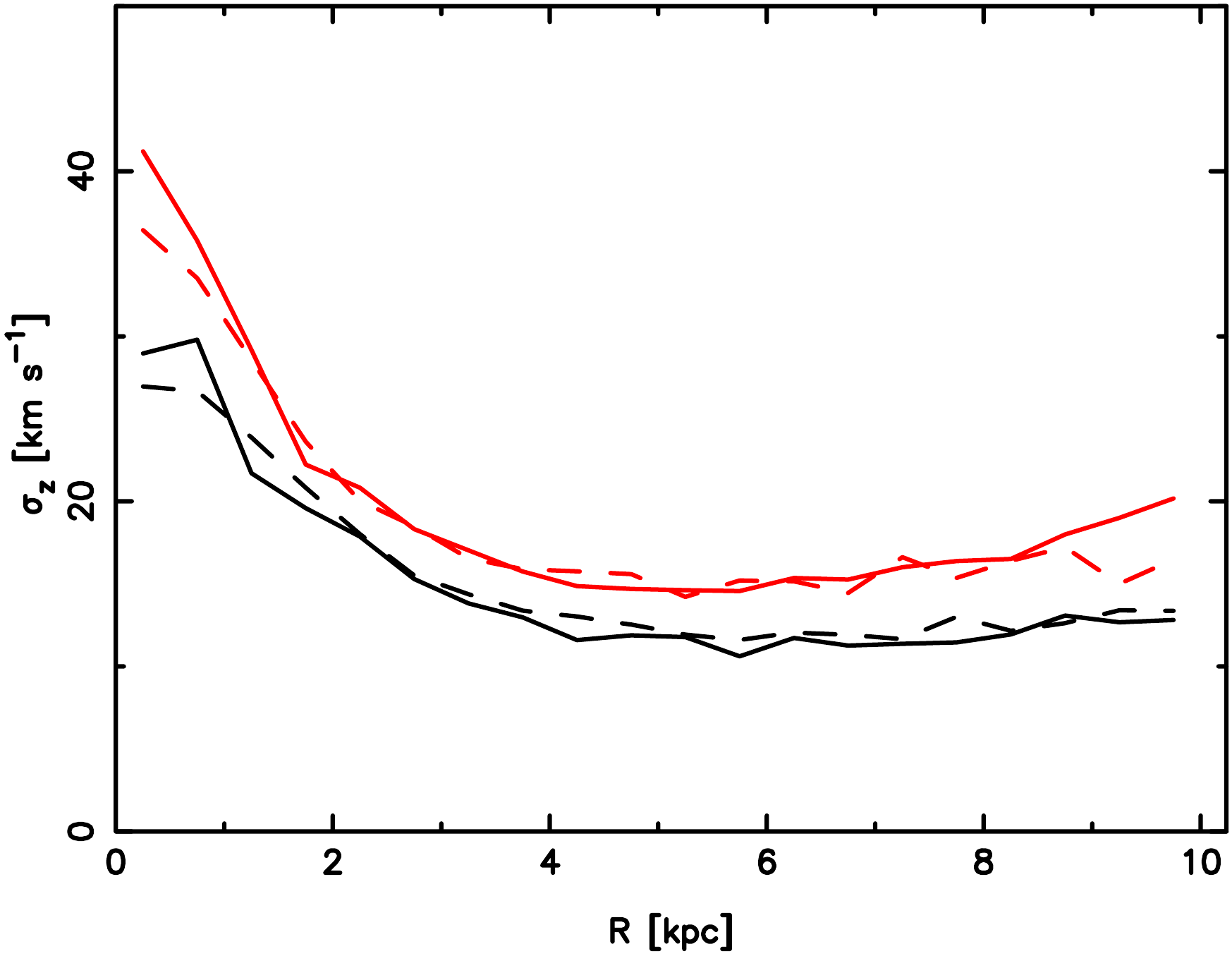}
\end{center}
\caption{The vertical velocity dispersion of stars, $\sigma_z$, at $z
  = 0.24$ (solid lines) and at $z = 0$ (dashed lines) for two
    models that have had a quiet history.  The black lines show model
  G14 while the red lines show model G15.}
\label{fig:vertdisps} 
\end{figure}

In order to verify that our simulation is not suffering from excess
artificial heating, we use the vertical velocity dispersion of stars,
$\sigma_z$, (since the in-plane velocity dispersions will increase
because of spiral structure).  We select two galaxies which have low
mass, which are not substantially tidally stripped and which never
formed bars, since bars vertically heat discs \citep[e.g.][]{Raha1991,
  Debattista2006}: galaxies G14 and G15.  These were replaced at $z = 0.25$;
Fig. \ref{fig:vertdisps} shows their vertical velocity dispersions
after relaxation, and at the end of the simulation after $\sim 3$ Gyr
of evolution.  Within 10~kpc $\sigma_z$ does not evolve substantially,
indicating that the models are not heating because of
artificial numerical effects.

\subsection{Formation of the cD galaxy}
\label{ssect:cd}

The stellar component of the cD galaxy in our simulation is comprised
of the stars from only a single galaxy that enters the cluster early
and settles to the centre.  At $z \simeq 1.65$, G01 enters the
cluster and undergoes several passes through the cluster centre before
finally settling there by $z\simeq 0.6$, thus forming a cD galaxy, at
the cluster centre.  The cD is a slowly rotating, pressure-supported,
spheroidal system, with a very extended envelope, which are all
properties typical of cD galaxies \citep{Oemler1976, Thuan1981,
  Schombert1986, Schombert1988, Oegerle1992, Bertola1995}.  If we had
included material that falls in earlier, when the cluster was more
chaotic, then the cD galaxy would have had a larger stellar mass.
However, this mass would have been more centrally concentrated than
that from G01 \citep{Diemand2005}, and would therefore contribute a
smaller mass fraction to the ICL.

The Fornax Cluster's cD galaxy, NGC~1399, has a stellar mass of
$2\times10^{11} ~\Msun$ within 5 kpc \citep{Saglia2000}, compared with
the cD in our simualtion, $4.4 \times 10^{10}~\Msun$.  The relatively
low mass cD galaxy in the simulation is a result of the replacements
starting from $z=1.65$; indeed at $z=2.7$ we identified an infalling
halo with a mass almost twice that of G01, which seems destined to 
fall onto the centre of the cluster and contribute to the cD.
Besides this, we identified a further 7 haloes falling into the
cluster between $z=3.8$ and $z=1.65$, with masses spanning $\sim
9\times 10^{9}~\Msun$ to $\sim 1.3 \times 10^{12}~\Msun$.  Thus a
further $5\times$ the mass of the halo of G01 is missed in our
replacements.  In the rest of this paper, we loosely
  refer to galaxy G01 as the cD galaxy, but it should be remembered
  that G01 is only a fraction of the total mass that would have gone
  into the cD galaxy of a real cluster like Fornax.


\section{Defining the ICL}
\label{sect:analysis}

\subsection{Imaging the ICL}

We analyse the results of the simulation at $z = 0$. To compare with
the Fornax Cluster, we view the simulation at a similar distance
($\sim$20 Mpc; \citealt{Blakeslee2009}) and assume a stellar
mass-to-light ratio of $M/L = 5 \Msun/\Lsun$ \citep{Rudick2006} to
convert the mass to a luminosity.  Though this choice is a
simplification, this is a characteristic value in the $V$-band for an
evolved stellar population at $z = 0$, which is the population from
which we expect the ICL to be comprised. This $M/L$ ratio is important
for when we define the ICL at fixed surface-brightness but, as
we show below, our preferred definition of the ICL depends on distance
from individual galaxies and is therefore independent of the assumed
$M/L$.

\begin{figure*}
\begin{tabular}{r}
  \includegraphics[width=110mm]{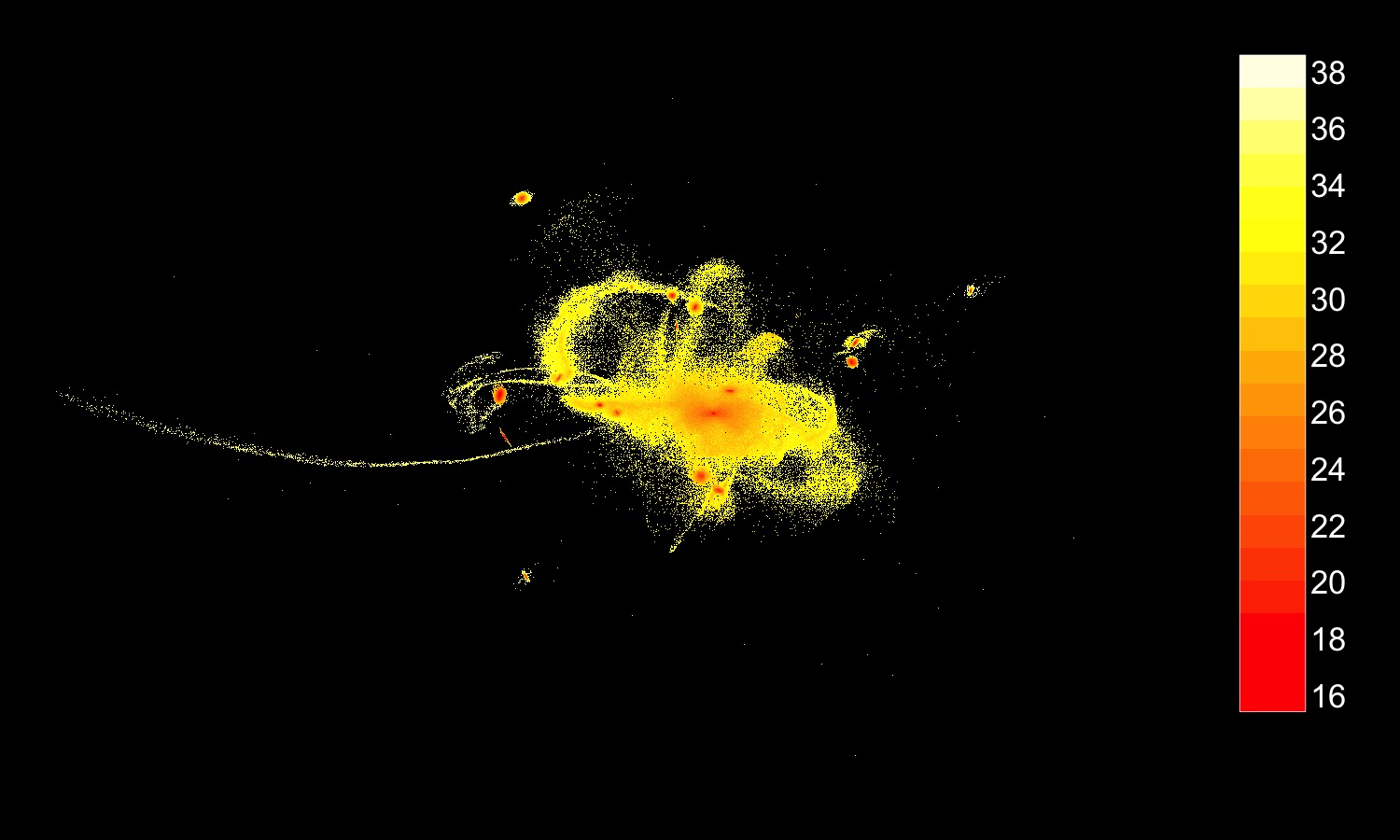} \\
  \includegraphics[width=110mm]{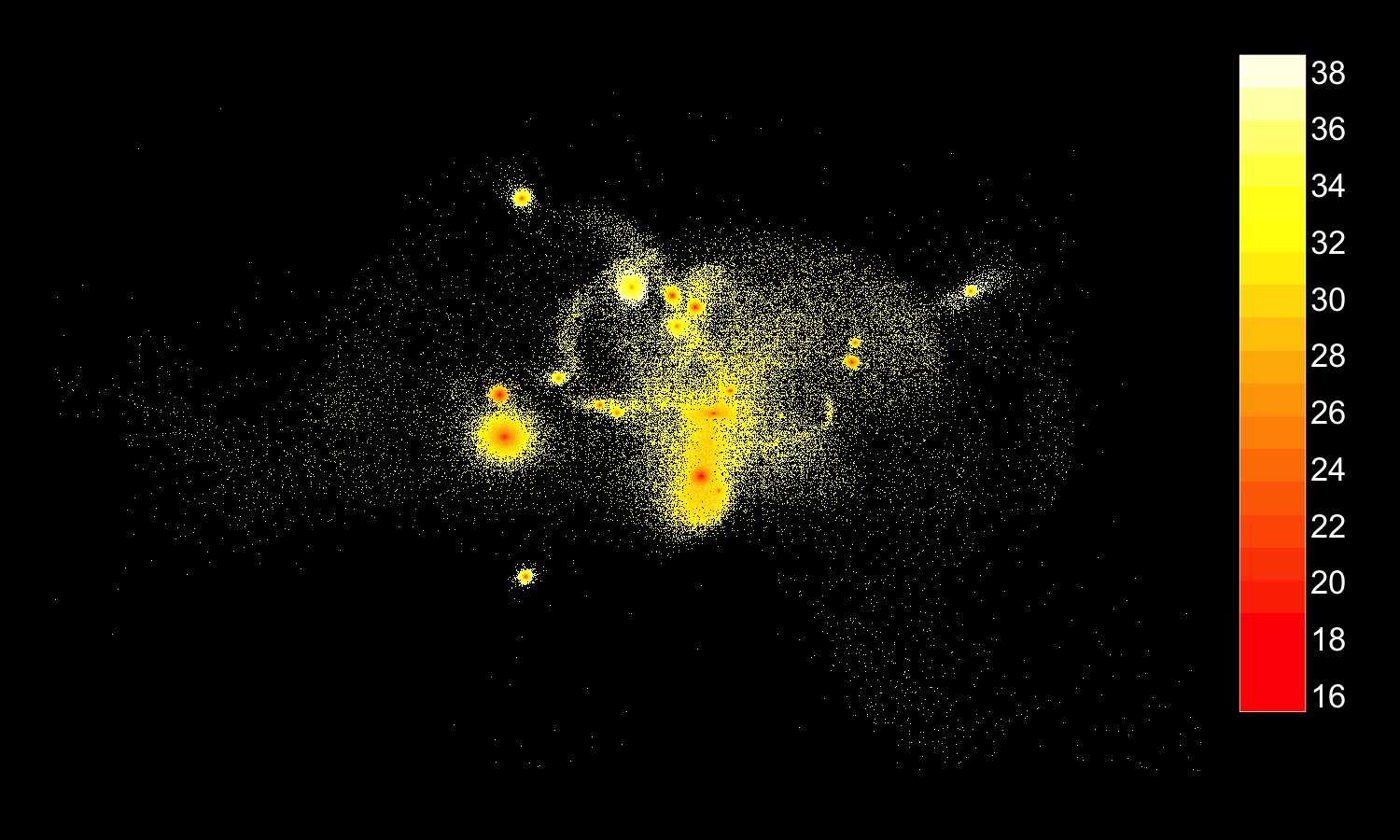} \\
\end{tabular}
\caption{The surface brightness (in the {\it V}-band) distribution of the disc (\textit{top}) and halo
  (\textit{bottom}) stars within the cluster.  Disc stars are more
  centrally concentrated, and exhibit more conspicuous narrow tidal
  features. }
\label{fig:contrib} 
\end{figure*}

We construct a series of cluster images measuring $1500 \times 1500$
pixels, with each pixel measuring $2 \times 2 ~\kpc$. This gives a
resolution of 0.023 arc-seconds per pixel compared with STIS imaging
on the {\it Hubble Space Telescope} which has 0.05 arc-seconds per
pixel. All of the images are centred on the cD galaxy.  We use three
orthogonal projections of the cluster along the $x$, $y$, and $z$ axes
of the simulation, averaging profiles over these three projections. We
treat the largest difference between the average and the individual
projections as our error estimate.  Fig. \ref{fig:contrib} shows the
distribution of all the disc (left) and halo (right) stars within the
cluster (including both those stars bound in galaxies, and those in the ICL). Stars which
originated in the discs are more centrally concentrated. The disc stars are also responsible for long, thin filaments
due to them being kinematically cooler populations which remain more
coherent after they are stripped from a galaxy \citep{Rudick2009}.

\subsection{Identifying the ICL}\label{sect:ICL}

To identify the ICL, past simulation works have used a
$V$-band surface brightness cut of 26.5 mag. $arcsec^{-2}$
\citep[e.g.][]{Rudick2006} (which we term the isophotal method), with
everything fainter than this limit being classified as ICL.  The
isophotal method is also the most commonly used definition of the ICL
for observational studies.  When applied to our simulation we found this
method excluded from the ICL dense areas of tidal streams.  Moreover
some of the fainter, low surface brightness galaxies were included
with the ICL.  Likewise \citet{Rudick2011} found that the isophotal
method produced less ICL compared to other definitions.

Other methods for identifying the ICL have been suggested, such as
binding energy (using halo detection algorithms to find the
  galaxy potentials and defining the ICL as stars outside these
  potentials; \citealt{Dolag2010}) and particle density threshold, which
requires knowledge of the number of luminous particles in a given
volume. (For more details on these methods, see \citealt{Rudick2011}.)
However these methods, while useful in simulations, are difficult to
apply to observations.

We therefore use a simple radial cut method to remove the galaxies,
taking 50 kpc around all the galaxies except galaxy G01 (the
cD), for which we use a cutoff of 100 kpc. These cuts are likely to remove any
bound contribution (see the appendix for evidence of this), leaving only the ICL.  While the choice of
cut may change the ICL fraction by $\sim50\%$, it does not change the
overall trends.  Our choice of 50~kpc is necessarily arbitrary, with
the main requirement being that very few stars still bound to a galaxy
are present beyond this radius.  We choose twice this
radius for galaxy G01 because cD galaxies usually have
extended stellar envelopes \citep[e.g.][]{Oemler1976,
  Schombert1986, Zhao2015}, with unclear boundaries
\citep[e.g.][]{Rudick2011}.  This will classify a smaller
  fraction of the stellar mass of the cluster as the ICL, but will be
more robust to misclassification of faint galaxies as ICL, while
being easy to implement observationally.  

Nonetheless, for
comparison, we have also carried out our analysis using a surface
brightness cutoff at 26.5 mag. $arcsec^{-2}$.  Our radial cut method
leads to roughly twice the amount of ICL as produced by a
surface brightness cut ($\sim 4\%$ using a radial cut versus $\sim
2\%$ using a surface brightness cut) and the fractional difference is
largest for disc stars, where fewer stars are classified as
  ICL by the surface brightness cut (by roughly a factor of 4,
whereas the difference is a factor of 3 for bulge stars).  In real
observations, a cut of 50 kpc would be too extreme if it was applied
to every galaxy, including the dwarfs, which we have not replaced.

\section{Results}
\label{sect:results}

\subsection{ICL Luminosity}
\label{ICL_L}

\subsubsection{ICL Fraction}
\label{sec:results_ICLfrac}

\begin{table*} 
\begin{centering} 
\caption{The values for each galaxy with the initial mass, mass lost to the ICL, the percentage of the ICL contributed from each galaxy (broken into disc and bulge stars). } 
\small{ 
\begin{tabular}{ccccccccc}
\hline 
 Galaxy & $z_i^a$ & Initial galaxy & Galaxy stellar & ICL mass & \% of mass & \% of ICL & \% of ICL & \% of ICL \\
        &         & stellar mass   &  mass at z=0   & at z=0   & lost to ICL &  & from bulge stars & from disc stars \\
     &          & ($10^{10} M_{\odot}$) & ($10^{10} M_{\odot}$)& $ (M_{\odot}$) &  & &  &  \\ \hline
  G01 & 1.65  & 4.79   & 4.43   & $3.6 \times 10^9$  &  8.14 & 21.9 & 0.53 & 21.4 \\
  G02 & 1.65  & 0.74   & 0.63   & $1.2 \times 10^9$  &  18.2 & 6.95 & 0.56 & 6.50  \\
  G03 & 1.13  & 1.30   & 1.09   & $2.1 \times 10^9$  &  19.2 & 12.7 & 0.83 & 11.86 \\
  G04 & 1.0   & 1.23   & 1.20   & $2.5 \times 10^8$  &  2.09 & 1.53 & 1.52 & 0.008 \\
  G05 & 1.0   & 1.13   & 1.03   & $1.0 \times 10^9$  &  10.2 & 6.31 & 3.00 & 3.31  \\
  G06 & 0.81  & 1.82   & 1.73   & $8.2 \times 10^8$  &  4.75 & 4.99 & 0.25 & 4.74  \\
  G07 & 0.81  & 0.18   & 0.18   & $3.3 \times 10^6$  &  0.18 & 0.02 & 0.02 & 0.0003  \\
  G08 & 0.81  & 5.69   & 5.00   & $6.9 \times 10^9$  &  13.8 & 41.75 & 39.1 & 2.67  \\
  G09 & 0.58  & 2.84   & 2.84   & $6.2 \times 10^6$  &  0.022 & 0.037 & 0.032 & 0.005 \\
  G10 & 0.4   & 0.57   & 0.567  & $1.4 \times 10^7$  &  0.25 & 0.086 & 0.018 & 0.068 \\
  G11 & 0.4   & 0.72   & 0.72   & $3.1 \times 10^6$  &  0.044 & 0.019 & 0.003 & 0.016 \\
  G12 & 0.4   & 0.39   & 0.38   & $6.4 \times 10^7$  &  1.66  & 0.39 & 0.06 & 0.32 \\ 
  G13 & 0.4   & 13.18  & 13.17  & $4.6 \times 10^7$  &  0.035 & 0.28 & 0.19 & 0.088 \\
  G14 & 0.25  & 0.27   & 0.27   & $6.5 \times 10^5$  &  0.024 & 0.004 & 0.0039 & 0.0 \\
  G15 & 0.25  & 0.16   & 0.16   & $7.6 \times 10^4$  &  0.0047 & 0.0005 & 0.0004 & 0.0001 \\
  G16 & 0.13  & 2.74   & 2.69   & $5.0 \times 10^8$  &  1.84 & 3.01 &  3.01 & 0.0  \\
  G17 & 0.13  & 0.05   & 0.05   & $3.4 \times 10^5$  &  0.065 & 0.002 & 0.002 & 0.0 \\
  G18 & 0.13  & 0.26   & 0.26   & $4.1 \times 10^5$  &  0.016 & 0.003 & 0.003 & 0.0 \\
  \hline
\end{tabular}
\begin{tablenotes}
        \item $^a$ Redshift at which the galaxy is inserted into the the cluster \\
\end{tablenotes}
}
\label{tab:ICL_frac} 
\end{centering} 
\end{table*}

Table \ref{tab:ICL_frac} lists the fraction of each galaxy which is
stripped from the galaxy and becomes ICL. In general the galaxies
which have contributed the greatest percentage of their stellar mass 
to the ICL are the ones that fell in earliest. Very little mass is lost to the ICL by
galaxies that fall into the cluster after $z = 0.6$, regardless of
stellar mass.

 Of the total stellar mass included in the $N$-body galaxy models
  ($3.81 \times 10^{11}$ \Msun), we classify ~4.4\% ($1.65 \times
  10^{10}$ \Msun) as ICL.  When calculating the fraction of the
  cluster in the ICL, we use the total cluster mass (i.e. including
  the mass still bound to other cluster members and the unbound mass)
  as the total mass. This definition has been used throughout this
  paper. Using the radial cut, $51\%$ of the ICL comes from disc
stars, while the rest is from haloes; if we exclude G01's (the central
galaxy) contribution to the ICL, the disc fraction is $38\%$ of the
remaining ICL.  These have been in the cluster longer, are more
centrally concentrated within the cluster and therefore have been more
disrupted than the younger galaxies.   It should be noted that the
  total stellar mass of the cluster as defined by an observer is
  likely to be greater than the mass described in this $N$-body
  simulation, as the simulation does not resimulate all the galaxies
  that fall into the cluster, nor a large fraction of the stars in the
  cD galaxy.  Since smaller galaxies and the cD would presumably
  contribute more to the total mass of the cluster than they would to
  the ICL, the ICL fraction would presumably decrease if they were
  included.

\subsubsection{Surface brightness distribution}
\label{sec:results_d+hstars}

Fig. \ref{fig:ICLonly} shows the distribution of the surface
brightnesses of the ICL and its contributions from the disc and halo
stars.  The stellar halo distribution peaks at a surface brightness
$\sim$ 27 mag. $~arcsec^{-2}$, while the disc stellar distribution
peaks at a brighter surface brightness of $\sim$ 25 mag.
$~arcsec^{-2}$.  It is clear from this figure that the isophotal cut method is too draconian, as it removes a significant number of originally disc stars from the ICL

\subsubsection{cD galaxy contribution} 
\label{sec:results_cD}

\begin{figure} 
\begin{center}
\includegraphics[width=0.35\textwidth]{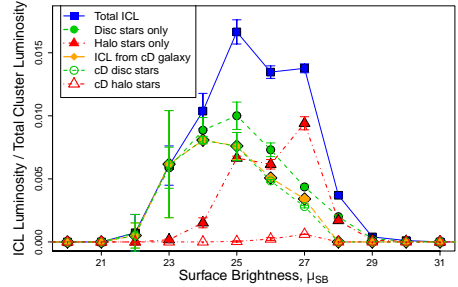}
\end{center}
\caption{The ICL luminosity fraction as a function of surface
  brightness (see the key for symbols.)  The change from a
  disc-dominated ICL to a halo-dominated ICL occurs at $\sim$ 26
  mag. $~arcsec^{-2}$ in the $V$-band.  Each pixel is $2\times 2
    \kpc$, corresponding to $0.023$ arcseconds per pixel at the
    distance of Fornax. }
\label{fig:ICLonly} 
\end{figure}

Table \ref{tab:cDpercents} presents the percentages that the galaxy
G01 (the cD) and its components contribute to both the total
cluster luminosity and to the ICL. G01 contributes $\sim$ 22\%
of the stars in the ICL (similar to the result in
\citealt{Murante2007}).  The majority of these stars originate from
the disc of G01.  Only 0.5\% of the ICL is made up of
halo stars from G01, even though for this galaxy the 
initial bulge to disc ratio is 0.12. The contribution from G01 to the ICL is shown in
Fig.  \ref{fig:ICLonly}.  As the orbit of G01 decays to the
centre, it makes multiple passes through the cluster centre and is
therefore strongly disrupted spreading the stars from the progenitor
over a large volume.

\begin{table*} 
\begin{center}
\caption{Percentages of the cluster luminosity contained in the ICL, and the ICL disc and halo stars from the central (cD) galaxy G01. }
{\scriptsize 
\begin{tabular}{c | c | c c c }
\hline
  Method                &  All ICL from G01 & ICL disc stars from G01   & ICL halo stars from G01 \\  \hline 
Contribution to total   & 0.95\% $\pm$ 0.21\%  & 0.92\% $\pm$ 0.20\%  & 0.03\% $\pm$ 0.012\%    \\ 
galaxy cluster          &   & & \\
Contribution to the     & 21.9\% $\pm$ 6.5\%   & 21.3\% $\pm$ 6.3\%   & 0.54\% $\pm$ 0.03\%  \\ 
ICL  &   & & \\ \hline
\label{tab:cDpercents} 
\end{tabular}
}
\end{center}
\end{table*}

\begin{figure} 
\includegraphics[width=0.35\textwidth,angle=0]{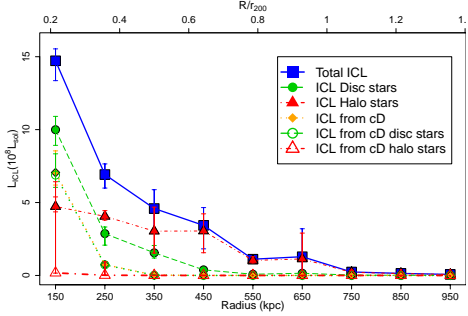}
\caption{The average ICL luminosity contribution as a function of
  radius for the different components.  The change from disc- to
  halo-dominated ICL occurs at $\sim$200 kpc. The contribution from
  the central (cD) galaxy also declines rapidly with radius and
  does not contribute any ICL beyond R $\sim$ 350 kpc.}
\label{fig:avlum} 
\end{figure}

\begin{figure} 
\includegraphics[width=0.35\textwidth,angle=0]{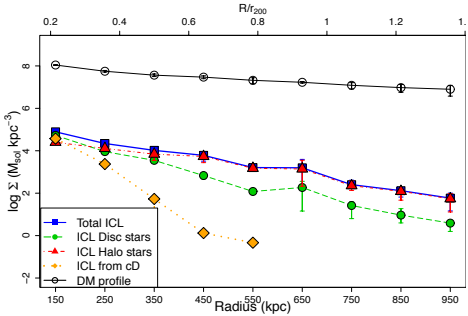}
\caption{The mass density profile as a function of cluster radius for
  the ICL, disc ICL stars, halo ICL stars, ICL from G01 (cD) and
  the dark matter profile.}
\label{fig:logslope} 
\end{figure}

\subsection{ICL Radial Profile}
\label{sect:radial}

Fig. \ref{fig:avlum} shows the radial profile of the ICL. Close to the
centre of the cluster, disc stars make up the greater part of the ICL,
dominated by those from G01 (the cD).  Beyond $\sim$200 kpc, the halo
stars become the main source of the ICL.  When the contribution of G01
is removed, at all radii the majority of the ICL is comprised of halo
stars.  This is because, further out in the cluster, only the less
gravitationally bound halo stars are stripped from galaxies.

Fig. \ref{fig:logslope} shows the profile of the mass density as a
function of radius for the ICL, ICL disc stars, and ICL halo stars,
compared with the dark matter density profile. The ICL stars have a
steeper density profile, including within $r_{\rm 200}/2$, than the
dark matter.  The steeper falloff of baryons relative to the dark
matter is expected, since the density profiles of stellar haloes
also tend to drop off more rapidly than those of the dark matter
haloes \citep[e.g.][]{Mandelbaum2010}.

\begin{figure} 
\begin{center}
\includegraphics[width=0.33\textwidth,angle=0]{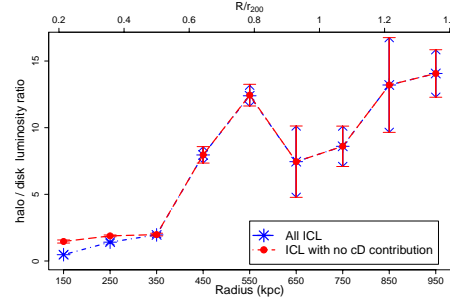}
\end{center}
\caption{The ratio of the halo-to-disc stars in the ICL as a function
  of radius from the centre of the cluster for the ICL as a whole (red
  circles, dashed line) and with the contribution from G01 (cD)
  removed (blue asterisks, dot-dashed line).}
\label{fig:HD_ratio} 
\end{figure}

Fig.  \ref{fig:HD_ratio} shows the ratio of the halo-to-disc stars for
the whole of the ICL, and with the contribution of G01 (cD)
  excluded.  G01 only has a minimal effect on the ICL halo
  component and contributes little ICL beyond a radius of $\sim$250
  kpc. At distances R $>$ 350 kpc, no stars from G01 (disc or
  halo) are present in the ICL.

\subsection{Stellar Ages}

\begin{figure} 
\includegraphics[width=0.35\textwidth,angle=0]{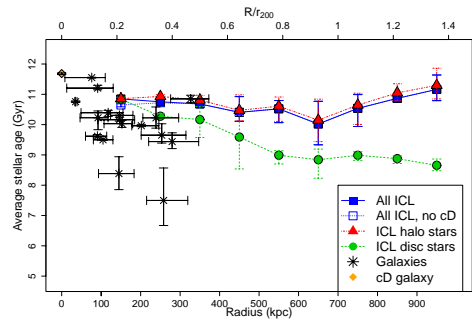}
\caption{The average age of disc and halo ICL stars as a function of
  radius.  (See the key for the symbols.) The plot also shows the ages
  with and without the contribution of G01 (the cD).  The disc
  stars are, on average, younger than the halo stars and the
  difference in ages increases at larger radii.  The average age of
  the galaxies is also indicated.}
\label{fig:ages} 
\end{figure}

The galaxies entered the cluster at different times. When the
galaxies fall into the cluster, we assume they cease star formation so
the stellar age depends on the initial age of the stars when the
galaxy enters the cluster. The semi-analytic models provide us with
stellar ages for the bulges and discs of each galaxy. Generally the
disc stars are younger than the halo stars.  Fig.  \ref{fig:ages}
shows the ages of halo ICL and disc ICL stars and the average age of
the ICL as a function of cluster radius.  Since halo ICL stars are
more prevalent at R $>$ 200 kpc (shown in Fig.  \ref{fig:avlum}), the
mean stellar age will be older further from the cluster centre.  Fig.
\ref{fig:ages} also shows the average age of the stars in the
galaxies. (The errors on the galaxy distance and stellar age are
derived from the variation in orientation.)  The galaxies themselves
exhibit a steeper average age profile than does the ICL; this is a
reflection of the early formation of the ICL and the later infall of
the cluster galaxies which remain at large radii.  As younger
galaxies lie further out from the cluster centre than the older
galaxies, they experience less disruptions and lose fewer stars to the
ICL.  Therefore the ICL consists mainly of stars removed from the
haloes of older cluster members.

\section{Discussion and Conclusions}
\label{sect:discussion}

We have presented a collisionless simulation studying the ICL in a
Fornax-cluster-mass system.  
For our simulated cluster, we estimate an ICL fraction of $\sim 4\%$ of
the cluster stellar mass.  This value is within the range of most of
the observations, though somewhat lower than simulated values
\citep[e.g.][]{Puchwein2010, Sommer2005}.  If we include G01
into the ICL, this fraction becomes $\sim 16\%$.  Table
\ref{tab:ICLfracs} compares the ICL fraction in our simulation with
that obtained by a number of observational and theoretical studies.

\begin{table*} 
\begin{center}
\caption{Comparison of the ICL fraction in our simulation and in
  observed, simulated and semi-analytic model clusters. } {\scriptsize
\begin{tabular}{c c c c c c }
\hline
      &  Cluster Mass  &  ICL fraction & ICL+BCG   & reference & ICL detection   \\  
      &   ($M_{\odot}$) &             &            &            & and notes   \\
      \hline 
observations &  $1.4\times10^{15}$ & 5.9$\pm$1.8\% & 8.2$\pm$2.5\%   & \citealt{Presotto2014} & Source subtraction $M(R < R_{500})$  \\
      & $>10^{15}$  &   & $>50\%$ & \citealt{Lin2004}  & Estimated via toy models \\
      & $10^{13}-10^{14}$ &  $2.69\pm1.6$ &    &  \citealt{McGee2010}  & Hostless SNe rates \\
      & $10^{13}-10^{14}$ &  $15.8\pm8$  &  & \citealt{Feldmeier2004} & Isophotal cutoff (non-cD clusters) \\
      & $\sim10^{14}$  &   1-4\%   &     & \citealt{Burke2012} & Isophotal cutoff (at $z\sim 1$) \\
      &      &  7-15\%   &   &  \citealt{Feldmeier2004b}  & Planetary nebulae (Virgo cluster) \\ \hline

simulations  & $10^{13}-10^{15}$ &   &  $\sim$45\% & \citealt{Puchwein2010} & Unbound particles$+$4 different    \\
\& SAMs      &                   &   &             &     & methods for separating ICL from BCG  \\

     & $10^{14}$      &  9-36\%   &       &  \citealt{Rudick2011} & Binding energy/double-Gaussian \\
     &                   &   &             &     & kinematic fit/density cutoff \\

      &  $10^{13}-10^{14}$  & 21-34\%  &   & \citealt{Sommer2005}  & Stars outside the tidal \\
    &                   &   &             &     &  radii of all galaxies \\
      & $>10^{13}-10^{15}$    &  $20-40\%$  &     &   \citealt{Contini2014}  &  SAM assuming stripped stars \\
      & $10^{14}-10^{15}$  & 20-30\%  &  & \citealt{Purcell2007} & SAM assuming stripped stars  \\
      & $4.1 \times 10^{13}$   & 4.3\%  & 16\% & This work\\  \hline
\label{tab:ICLfracs} 
\end{tabular}
}
\end{center}
\end{table*}

In this mass
regime, the majority of the ICL outside G01 (the cD galaxy) is
derived from the haloes of galaxies: 
$\sim 51\%$ of its mass from stars that started their life in galactic
discs ($\sim 38\%$ if G01 is excluded).
Since our starting models have extended pure exponential density
profiles to large radii, whereas the majority of real disc galaxies
are truncated, the amount of stellar mass in the simulation's outer
discs is overestimated.  This gives us confidence that our
disc ICL fraction is an upper limit.  
 However, inside $0.25r_{200}$, the ICL is dominated by stars from the disc of G01,
the galaxy that becomes part of the cD (see
Fig. \ref{fig:avlum}).

We studied the contribution to the ICL from individual galaxies and
found that the largest, oldest galaxies contribute the most to the
ICL. This agrees with \citet{Contini2014} who found 26\% of ICL comes
from galaxies with stellar masses between $10^{10.75}$ and
$10^{11.25} ~M_{\odot}$ and 68\% of ICL is from galaxies with $M_{*}
>10^{10.5} ~M_{\odot}$. For galaxies of mass $M_* <4\times 10^9 ~\Msun$,
their contribution to the total ICL is negligible.  We note that this is
close to our cutoff for the replaced galaxies.  Thus it is reasonable
to suppose that we have not missed a significant fraction of the ICL
from low mass galaxies in our simulation.

The radial distribution of the ICL shows a change in composition from
being disc-dominated near the cluster centre to stellar halo-dominated
by $\sim 25\%$ of $r_{200}$ (Fig. \ref{fig:avlum}). This
change is due to G01 contributing the majority of the disc ICL
stars. The distance is approximately half the distance found by
\citealt{Murante2007} for a cluster of virial mass $1.6-2.9 \times
10^{14} h^{-1} ~\Msun$.

The stellar ages are different between the galaxies and the ICL as a
function of radius. The older galaxies within the cluster tend to lie
near the cluster centre while the younger galaxies lie further out,
which reflects the time of infall into the cluster.  ICL
stars are, on average, older than the average galaxy stars,
particularly at large radii, with the average age nearly constant or
increasing slightly with radius. This is similar to the result found
by \citet{Puchwein2010}.  The radially nearly constant age of the
ICL stars is due to most of them having been stripped from galaxies
that fell in to the cluster at an early epoch.  The younger galaxies are
not as heavily stripped so do not have as large an impact on the ICL.
This old ICL is different from the semi-analytic model result of
\citet{Contini2014}, who find a predominantly younger ICL (forming
since $z=1$).

Our simulation does not include gas; nonetheless we produce an ICL
fraction comparable to that observed \citep[e.g.][]{Burke2012,
  Feldmeier2004b}.  Hydrodynamical simulations instead have produced
much higher ICL fractions \citep[e.g.][]{Puchwein2010,
  Sommer2005,Cui2014}, some of it from stars forming directly out of
cluster gas \citep{Puchwein2010}.  On the other hand, the ages of
galaxies in our simulation are similar to those of
\citet{Puchwein2010}.  The main assumption upon which the ages of our
galaxies rest, that star formation is quenched upon the galaxies
entering the cluster, is borne out by the fully cosmological
simulations.

In our simulation, the ICL density profile falls off more rapidly than
that of the DM.  That the baryons are more centrally concentrated than
the dark matter is not surprising since they lose energy in settling
to the centre of haloes \citep{White1978}.

\section{Acknowledgements} 

KAH would like to thank the University of Central Lancashire for the
grant to start this work, and Virginia Tech for allowing its
completion, also John Feldmeier for his discussion.  VPD is supported
by STFC Consolidated grant \#ST/J001341/1.  VPD, AJC and BBT would
like to acknowledge the Undergraduate Research Internship Scheme
(2011) for the funding to support early analysis of this simulation,
and also the Nuffield Foundation (2011) undergraduate bursary scheme
and the Royal Astronomical Society (2012) for the support that
contributed to the initial analysis in this paper.  EWP acknowledges
support from the National Natural Science Foundation of China under
Grant No. 11173003, and from the Strategic Priority Research Program,
``The Emergence of Cosmological Structures'', of the Chinese Academy
of Sciences, Grant No. XDB09000105.  The authors acknowledge the Texas
Advanced Computing Center (TACC)\footnote{http://www.tacc.utexas.edu}
at The University of Texas at Austin for providing HPC resources that
have contributed to the research results reported within this paper.
We thank Sam Earp for preparing Figure \ref{fig:massmass} for us.

%
%
 \bsp
\label{lastpage}

\appendix
\section{Density profiles, Dark matter volume profiles and orbits of the inserted model galaxies}\label{Appendix 1}

\begin{figure*}
\begin{tabular}{cc}
     \includegraphics[width=0.4\textwidth,angle=-90]{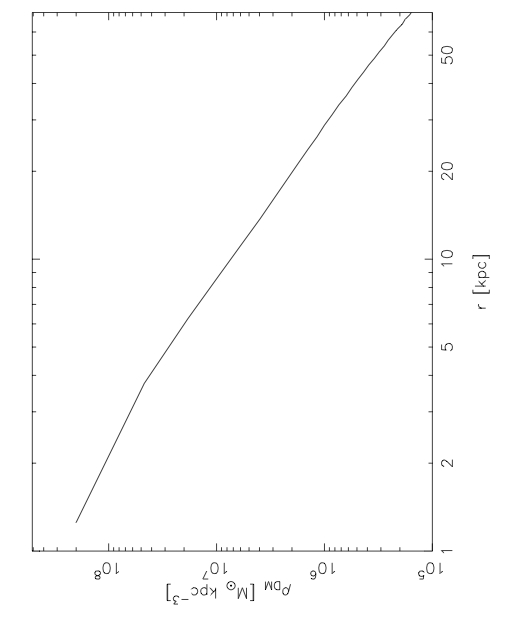} 
  &  \includegraphics[width=0.4\textwidth,angle=-90]{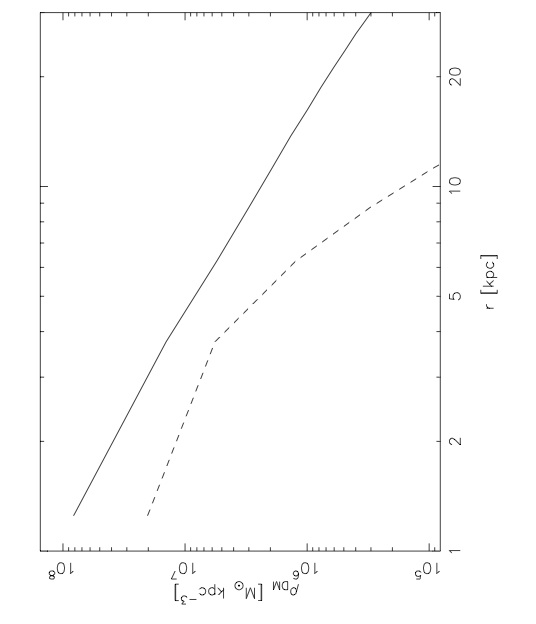} \\
     \includegraphics[width=0.4\textwidth,angle=-90]{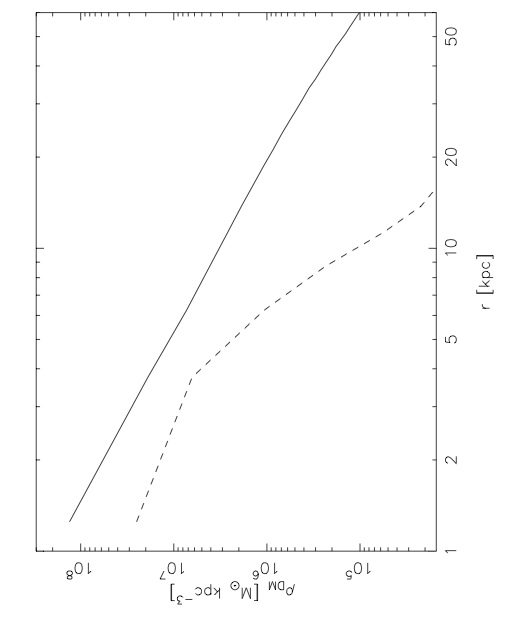} 
  &  \includegraphics[width=0.4\textwidth,angle=-90]{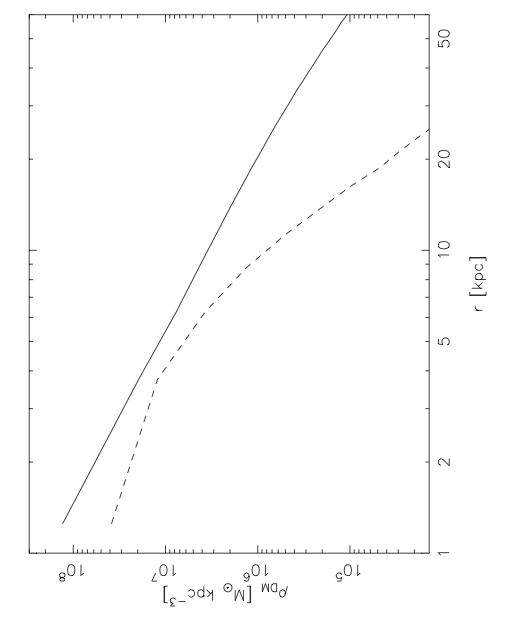} \\
     \includegraphics[width=0.4\textwidth,angle=-90]{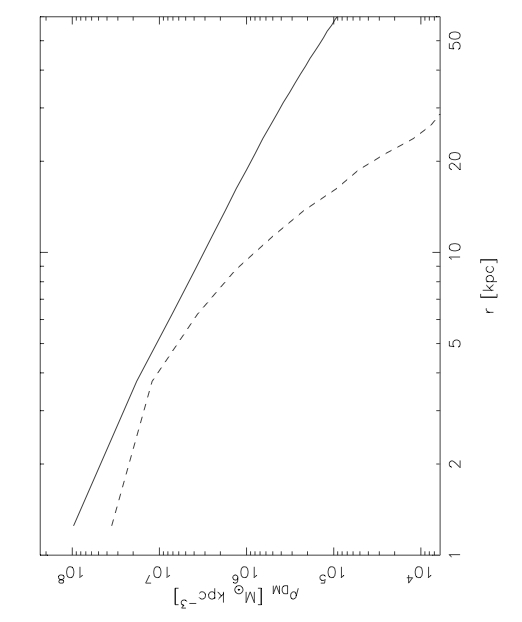} 
  &  \includegraphics[width=0.4\textwidth,angle=-90]{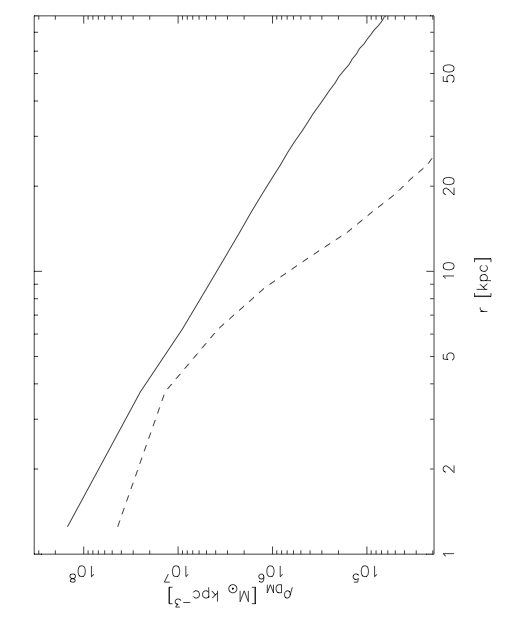}  \\
\end{tabular}
\caption{ Dark matter volume density profile for G01 - G06. The solid line is the
halo at the time of insertion, and the dashed line is at $z = 0$. No $z=0$ dark matter 
profile is shown for G01 because the galaxy is now at the centre of the cluster. 
The figures for the other galaxies can be found online. }
\label{fig:DM_1} 
\end{figure*}

\begin{figure*}
\begin{tabular}{cc}
     \includegraphics[width=0.35\textwidth,angle=0]{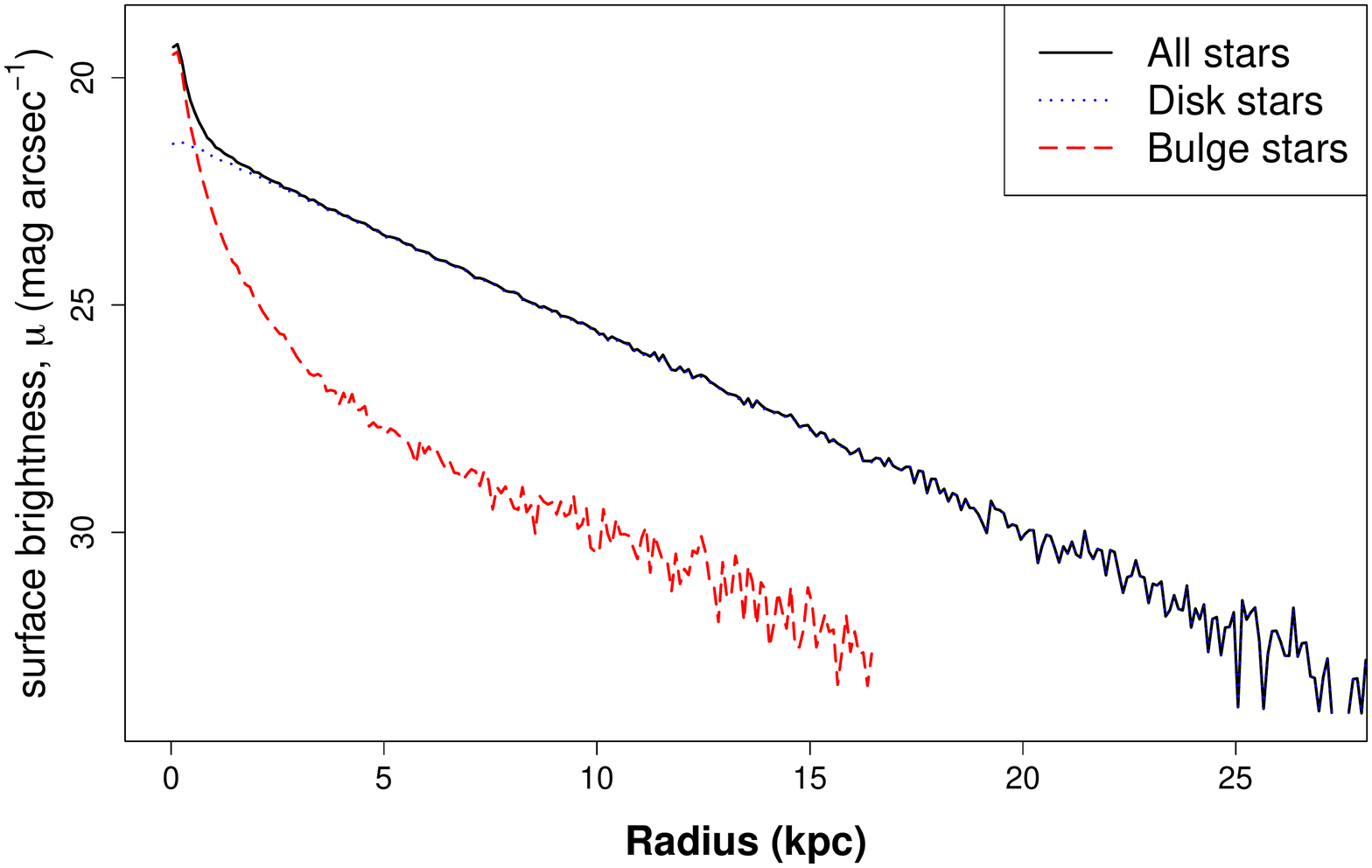} 
  &  \includegraphics[width=0.35\textwidth,angle=0]{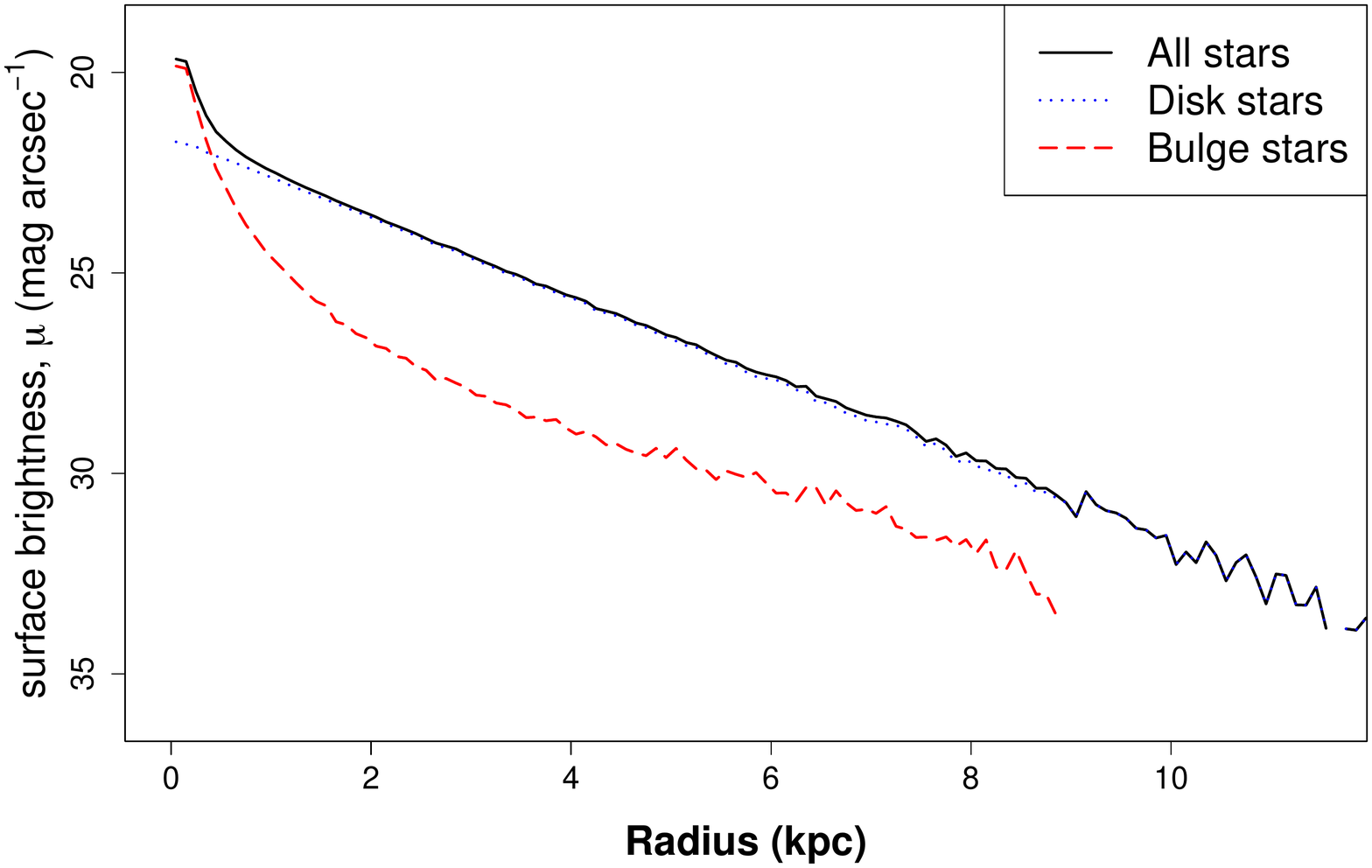} \\
     \includegraphics[width=0.35\textwidth,angle=0]{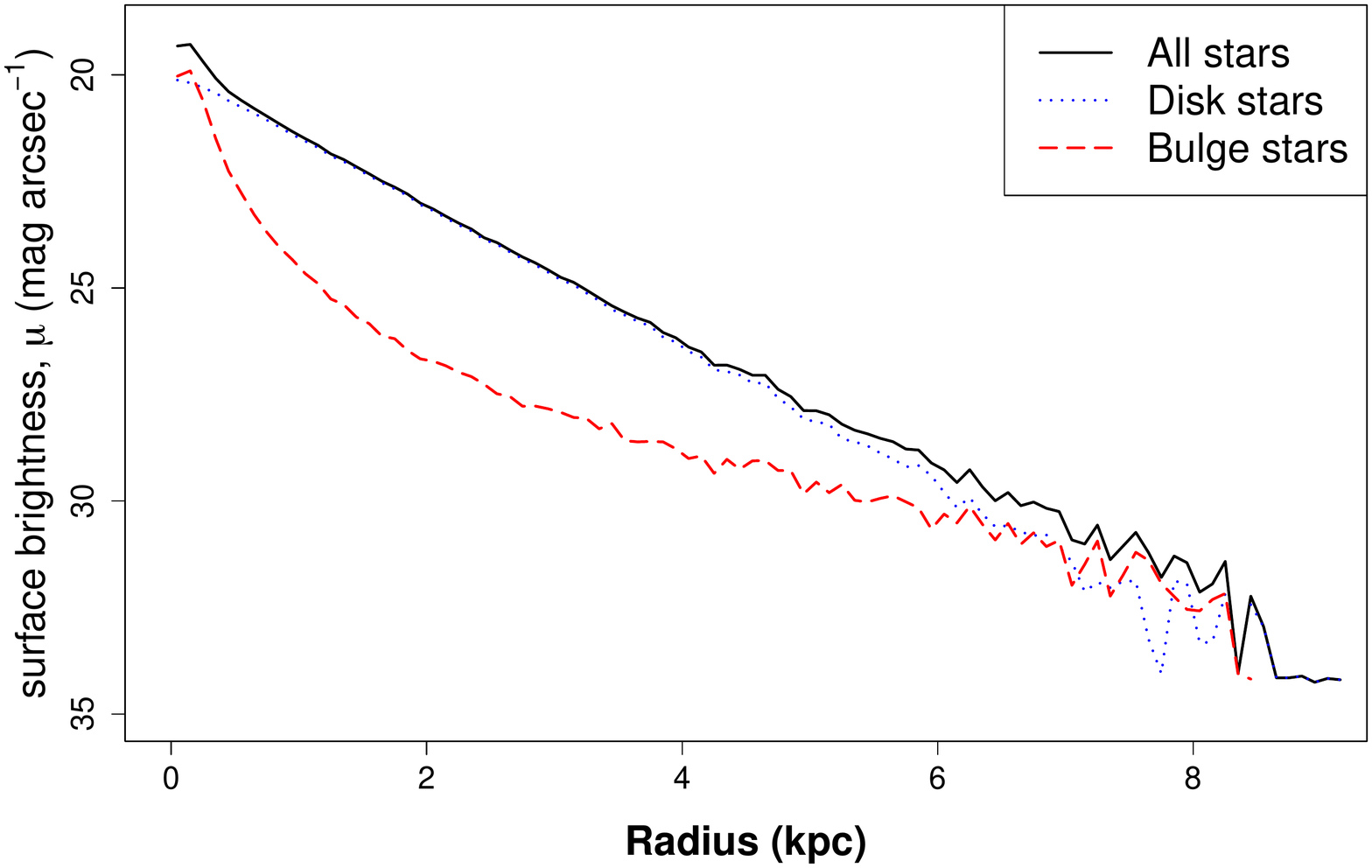} 
  &  \includegraphics[width=0.35\textwidth,angle=0]{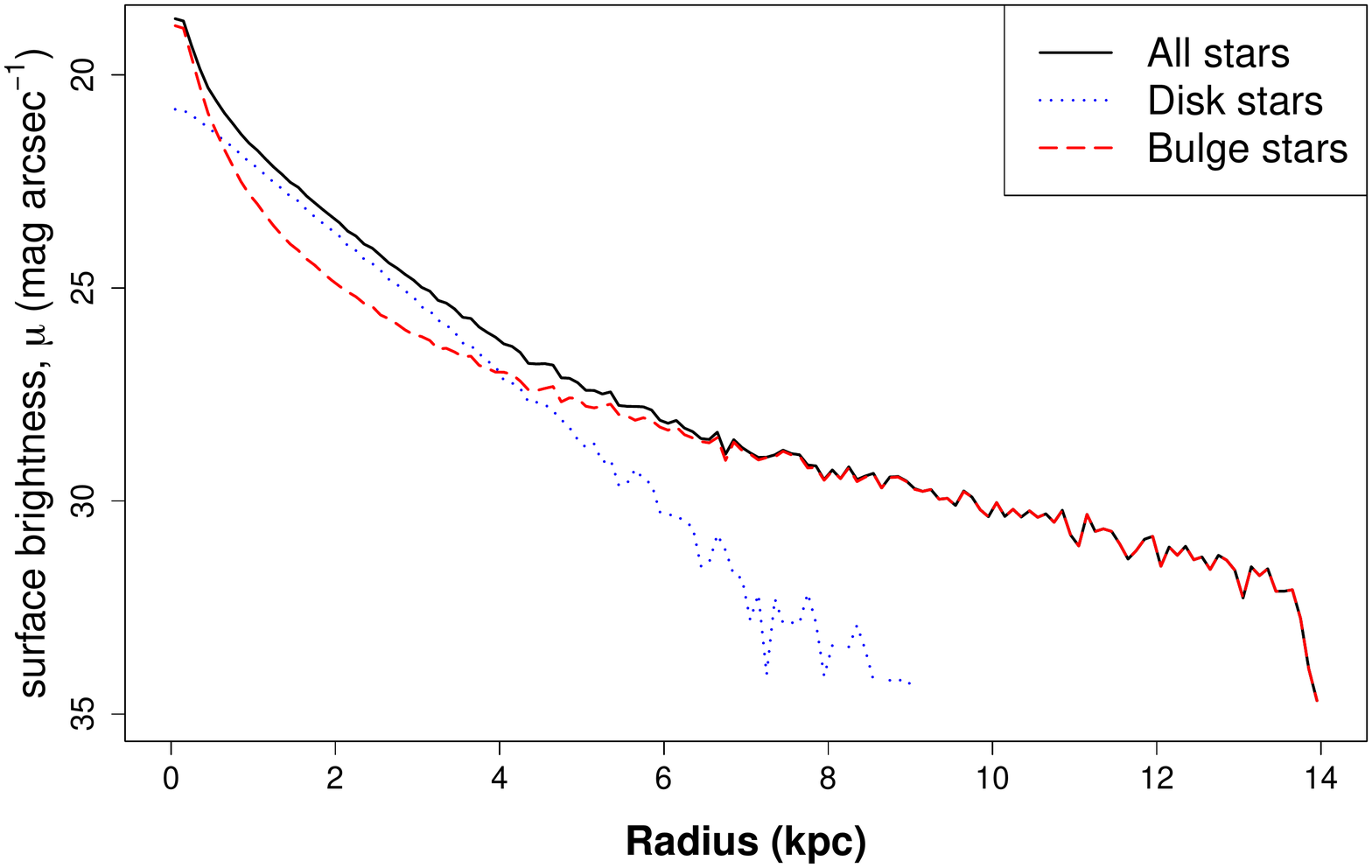} \\
     \includegraphics[width=0.35\textwidth,angle=0]{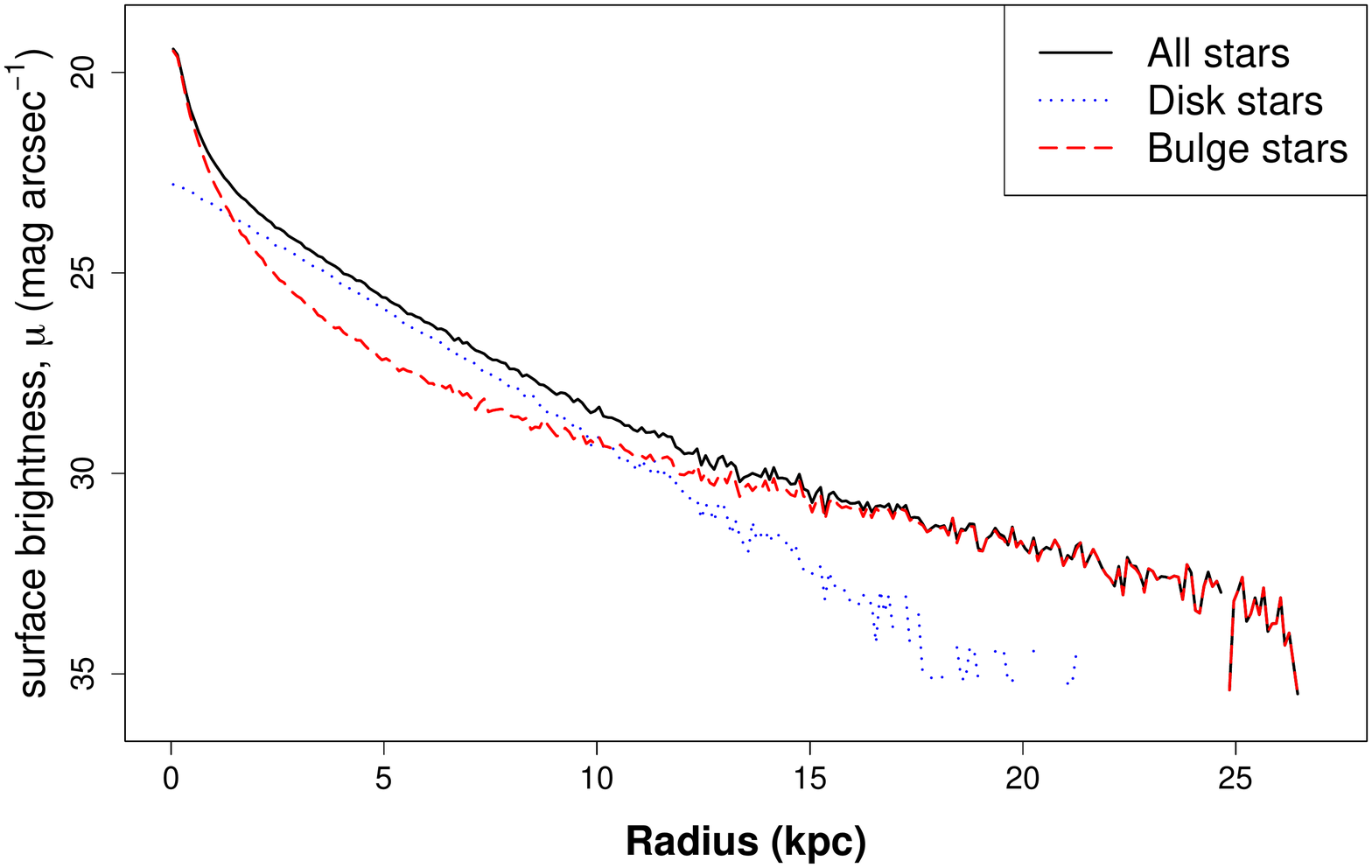} 
  &  \includegraphics[width=0.35\textwidth,angle=0]{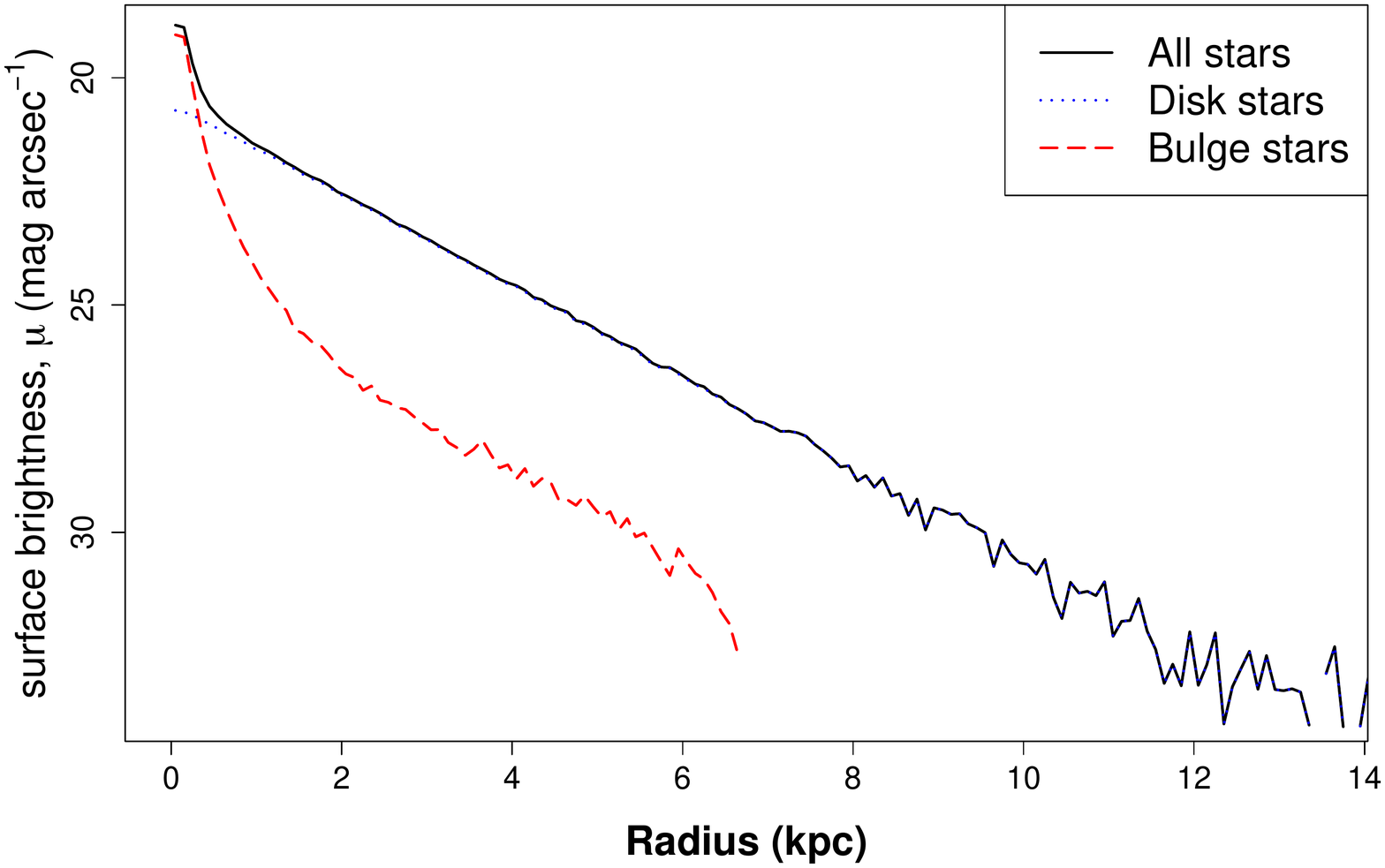} \\
\end{tabular}
\caption{ Surface brightness profile in the $V$-band for the whole galaxy (black solid), 
the disc stars (blue dotted) and the bulge stars (red dashed) when the galaxy is first
inserted into the simulation for G01 - G06. 
The figures for the other galaxies can be found online. }
\label{fig:SB_1} 
\end{figure*}

\begin{figure*}
\begin{tabular}{cc}
     \includegraphics[width=0.3\textwidth,angle=90]{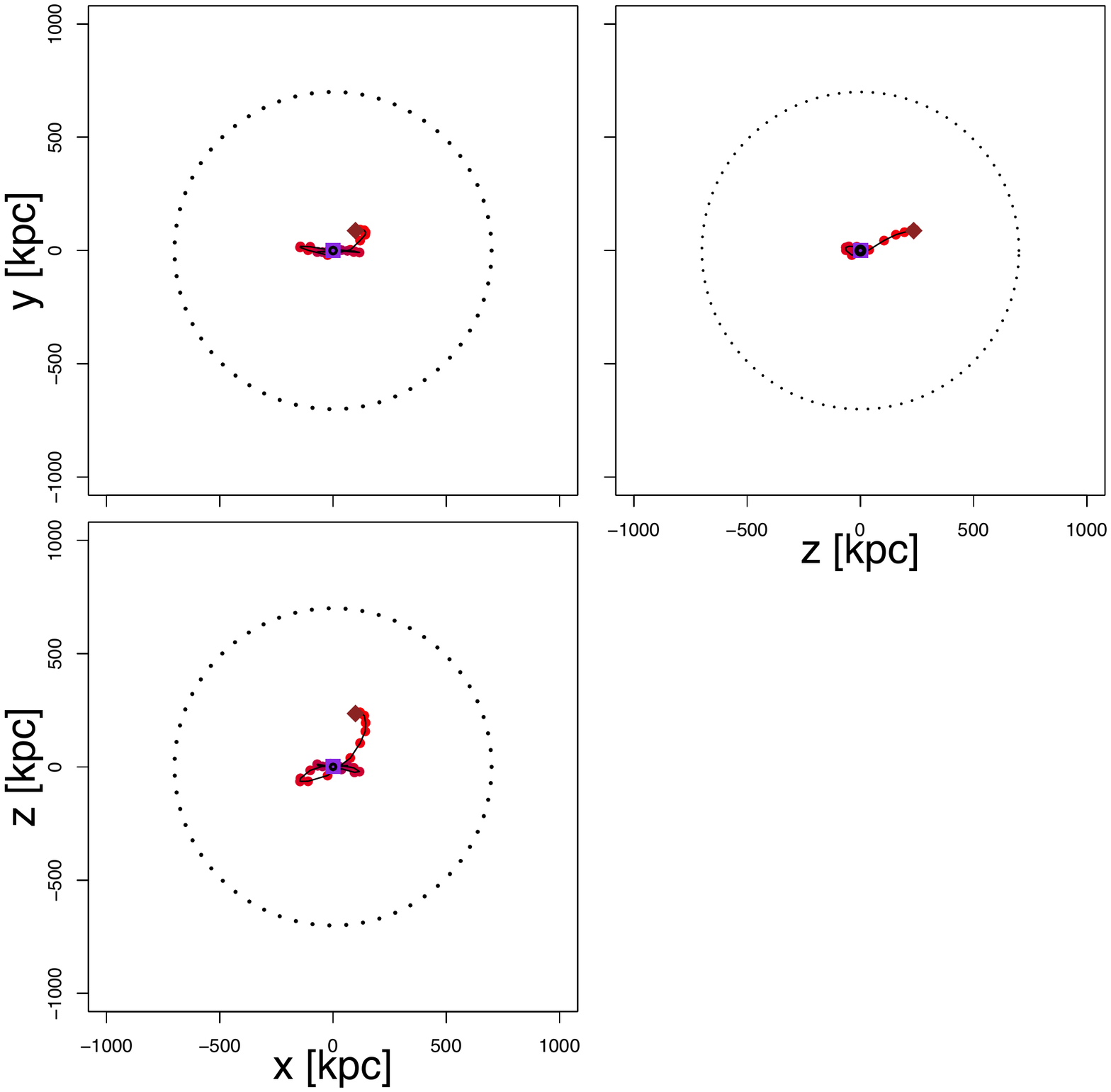} 
  &  \includegraphics[width=0.3\textwidth,angle=90]{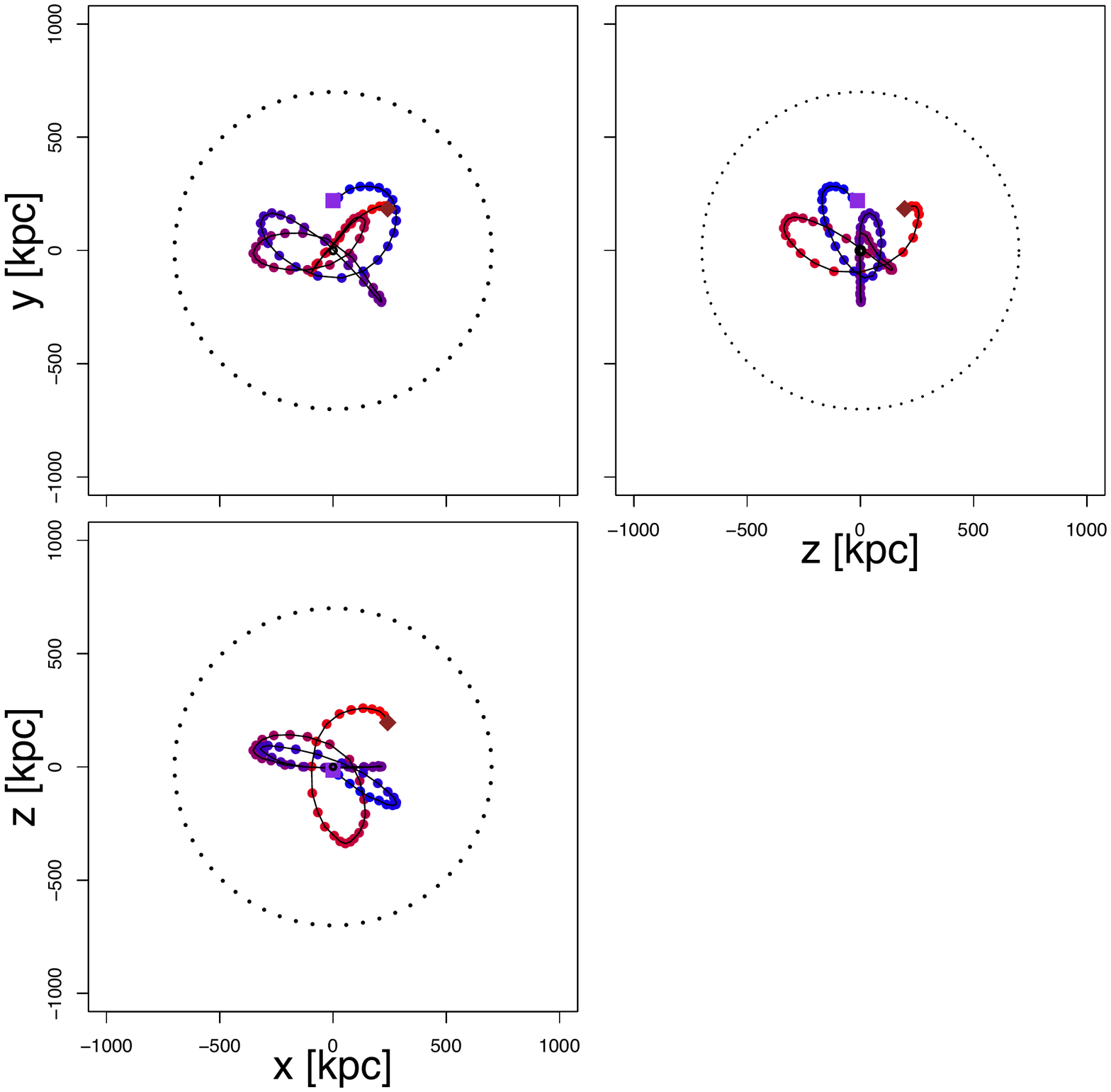} \\
     \includegraphics[width=0.3\textwidth,angle=90]{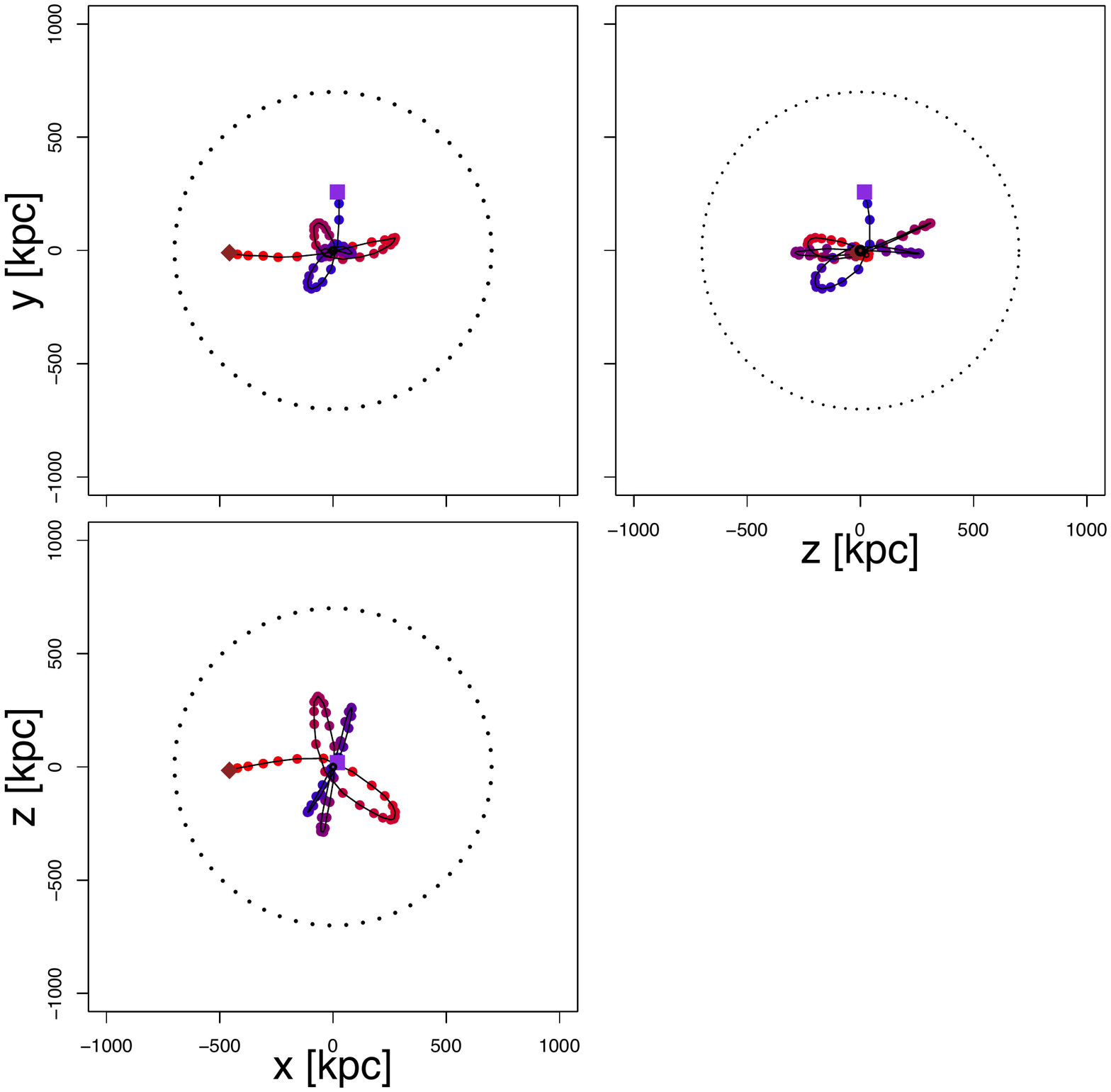} 
  &  \includegraphics[width=0.3\textwidth,angle=90]{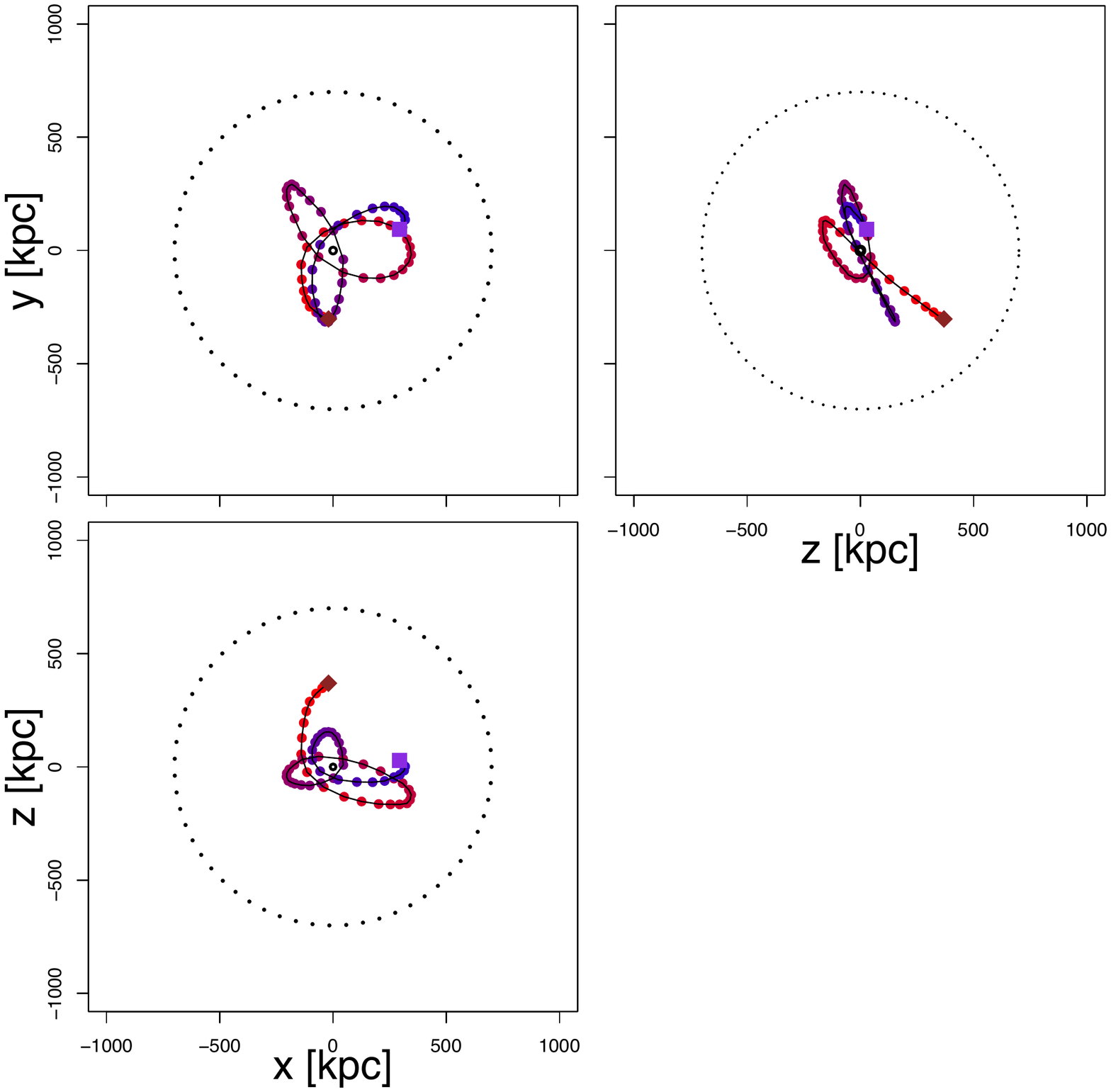} \\
     \includegraphics[width=0.3\textwidth,angle=90]{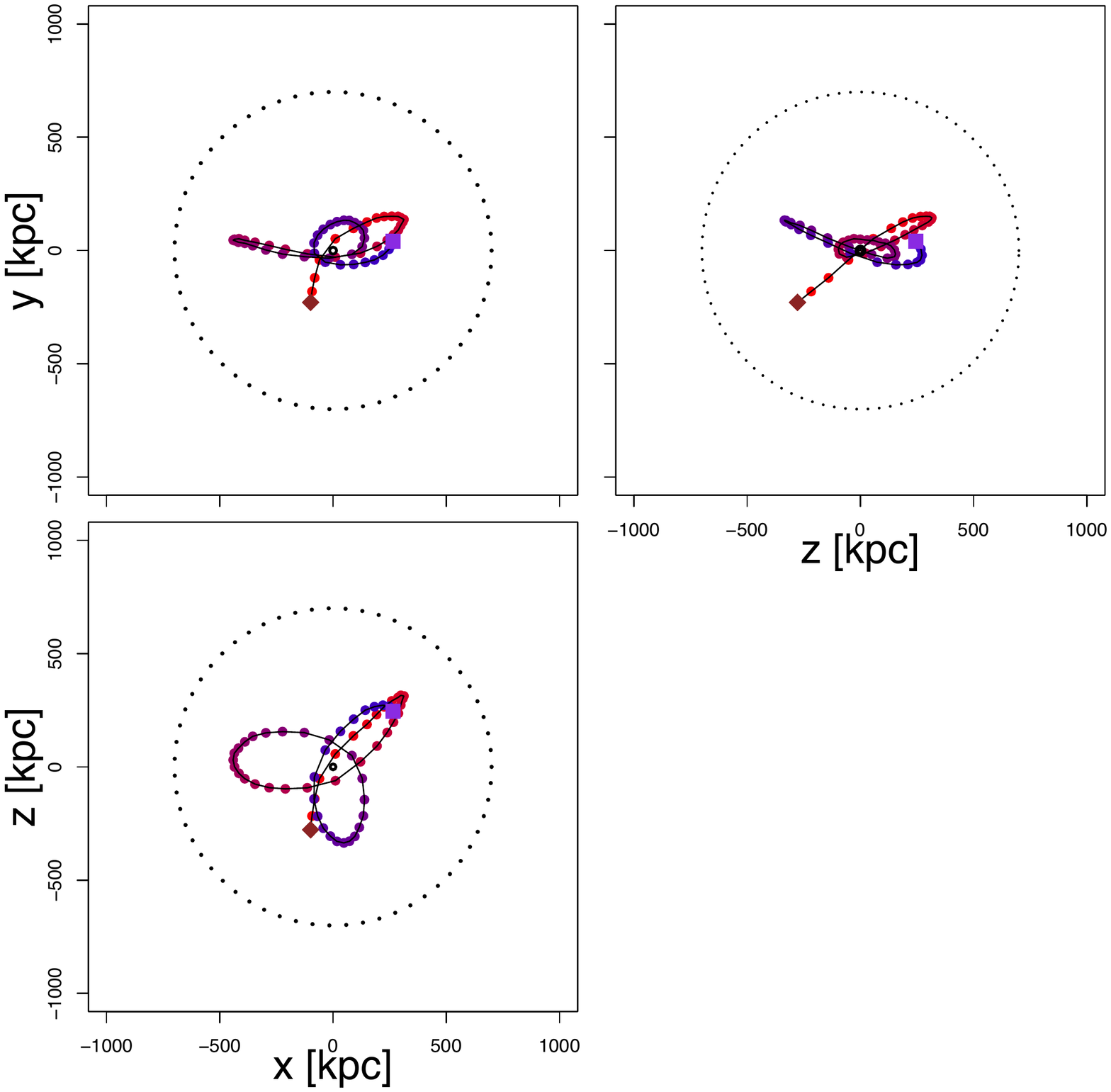} 
  &  \includegraphics[width=0.3\textwidth,angle=90]{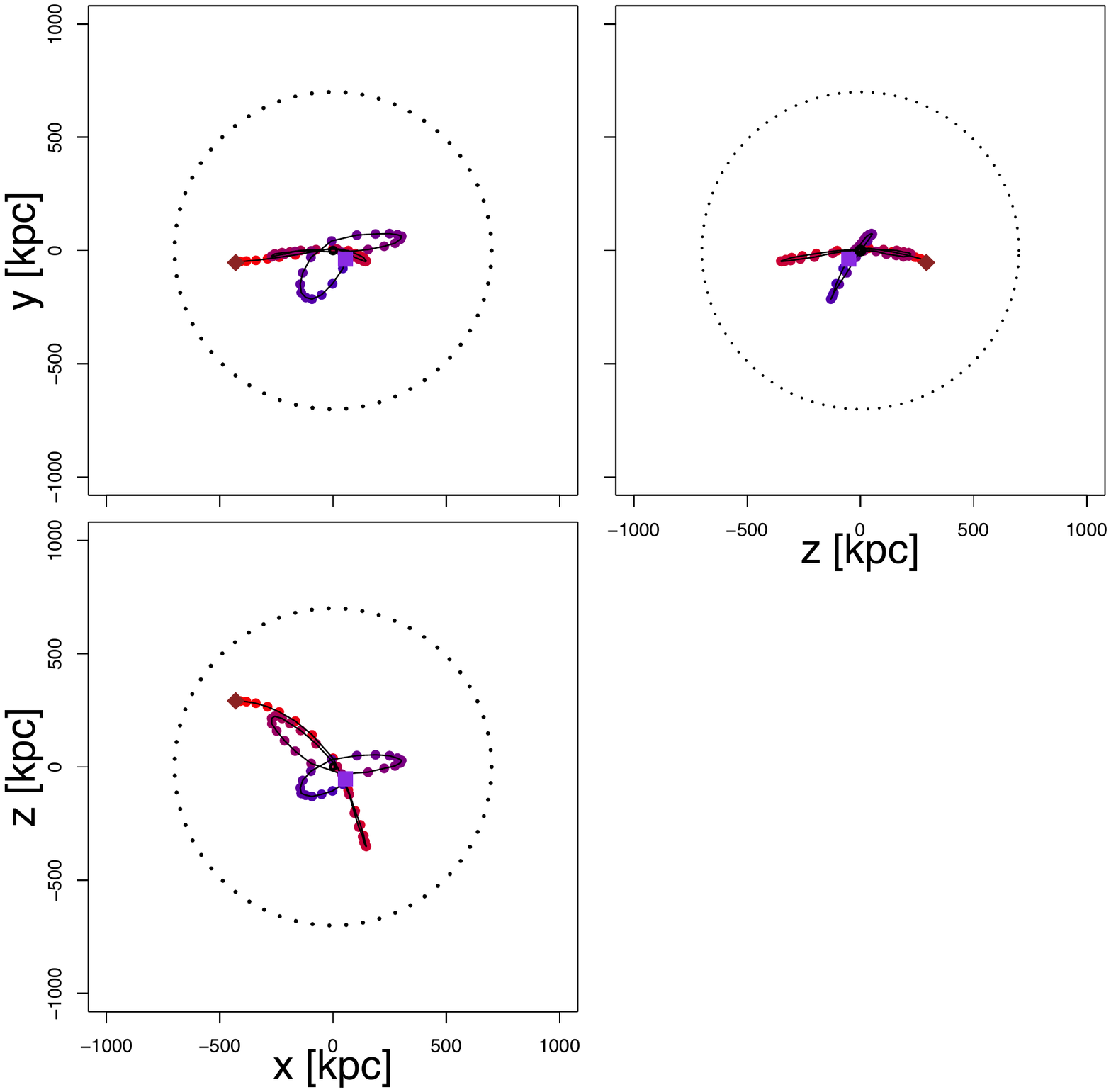} \\
\end{tabular}
\caption{ The orbit taken by galaxies G01 - G06. The
  initial insertion point is given by a brown diamond. The path moves
  from red to blue ending in the violet square which indicates the
  final position at $z = 0$.  The dotted black circle shows the virial
  radius, and the central blue circle shows the half mass radius of
  model G01, both at $z=0$. The figures for the other galaxies can be found online.}
\label{fig:orbit_1} 
\end{figure*}

 Fig. \ref{fig:DM_1}  shows the dark matter volume density
profile for all the models.  The solid line is the halo at the time of
insertion, and the dashed line is at $z = 0$.  There is no profile for
G01 at z=0 since the halo is that of the cluster.  For the rest
  of the galaxies, refer to the online figures.  
  Fig. \ref{fig:SB_1} shows the surface brightness profile in the
$V$-band for the whole galaxy (black solid), the disc stars (blue
dotted) and the bulge stars (red dashed) when the galaxy is first
inserted into the simulation.  For the rest of the galaxies,
  refer to the online figures.  Fig. \ref{fig:orbit_1} shows
the orbit taken by that galaxy, with the galaxies moving from red to
blue as time progresses. The initial insertion point is given by a
brown diamond. The path moves from red to blue ending in the violet
square which indicates the final position at $z = 0$. The dotted
circle shows the cluster virial radius while the smaller solid circle
in the centre shows the half light radius of the cD galaxy at $z = 0$.
For the rest of the galaxies, refer to the online figures.

\newpage

\end{document}